%% file: main_arXiv.tex
\documentclass[preprint,3p,12pt]{elsarticle}
\biboptions{authoryear}
\usepackage{amssymb}
\usepackage{amsmath}
\usepackage[dvipsnames]{xcolor}
\usepackage{dynamo_symbols}
\usepackage{hyperref}
\usepackage{xr}
\usepackage{overpic}
\usepackage{lineno}
\usepackage{gensymb}
\usepackage{comment}
\journal{Earth Planet. Sci. Lett.}

\begin{document}

\begin{frontmatter}
\title{Paleomagnetic signatures of core-mantle interactions inferred from top-heavy thermochemical geodynamo simulations}
\author[aff1,aff2]{Souvik Naskar\corref{cor1}} 
\ead{sn728@cam.ac.uk} 

\author[aff1]{Jonathan E. Mound} 
\author[aff1]{Christopher J. Davies} 
\author[aff1]{Hannah F. Rogers} 
\author[aff1]{Stephen J. Mason} 
\author[aff1]{Andrew T. Clarke} 

\cortext[cor1]{Corresponding author}
\affiliation[aff1]{organization={School of Earth and Environment},
            addressline={University of Leeds}, 
            city={Leeds},
            postcode={LS2 9JT}, 
            state={},
            country={UK}}
\affiliation[aff2]{organization={Department of Applied Mathematics and Theoretical Physics},
            addressline={University of Cambridge}, 
            postcode={CB3 0WA}, 
            country={UK}}            

\begin{abstract}
The time-averaged geomagnetic field provides crucial insights into deep Earth dynamics and thermal core-mantle interactions. Paleomagnetic observations and numerical dynamo simulations are equivocal regarding the longitudinal structure of the time-averaged field, though the latter have often considered a generic buoyancy source, which may obscure distinct signatures of thermal and chemical buoyancy that arise near the equator and poles, respectively. In this study, we present a new suite of top-heavy geodynamo simulations, varying the relative strengths of thermal and chemical driving and comparing the resultant magnetic signatures to observational field models spanning centuries to tens of thousands of years.
None of the spatially-averaged measures of field morphology and variability we tested could robustly distinguish between different levels of chemical driving or the presence of heterogeneous outer boundary heat flux. On the other hand, observational constraints requiring longitudinal variations in time-averaged inclination anomaly are readily matched by simulations with heterogeneous outer boundary thermal forcing, in contrast to those with homogeneous mantle heat flux. 
Longitudinal field structures are reduced, but not erased, by elevated chemical driving, which also promotes the formation and deepening of polar minima in the radial magnetic field.
Our simulations indicate that both the strong heat flux heterogeneity and chemical driving in Earth's core are likely to result in small but persistent departures from the geocentric axial dipole approximation.

\end{abstract}







\end{frontmatter}



\section{Introduction}
Earth’s magnetic field has been sustained for billions of years by dynamo action, powered by secular cooling of the liquid core as heat slowly escapes to the mantle. Although mantle convection operates on much larger spatial and temporal scales than core convection, heterogeneities in the lowermost mantle may exert a first-order influence on core dynamics and the geodynamo. Seismic anomalies in the lowermost mantle reflect thermal and chemical variations that result in substantial lateral variations in the heat flux across the core-mantle boundary \citep[CMB; ][]{masters_1996, stackhouse2015first, dannberg2024changes,deschamps_2026}, which should influence core convection and the geomagnetic field \citep[e.g.][]{bloxham_2000, olson_2002, gubbins_2007a, amit_2015a}. However, the signature of lowermost mantle heat flow heterogeneity in geomagnetic field behaviour and the expression of this signature in available geomagnetic and paleomagnetic observations has remained uncertain and has been heavily debated \citep{gubbins_2003, amit_2015b, olson_2016}. This issue can now be tackled anew by combining advances in global time-dependent geomagnetic field models \citep[e.g.,][]{panovska_2019} and paleomagnetic data compilations \citep[e.g.,][]{cromwell_2018,tauxe_2024} with targeted suites of numerical geodynamo simulations \citep[e.g.,][]{biggin_2026}.  

Previous geodynamo simulations have shown that imposed thermal CMB heterogeneity can influence both the long-term morphology and the stability of the field that is generated. Morphological characteristics of thermal CMB heterogeneity include the persistence of concentrated patches of magnetic flux in the time-averaged field (TAF) at high \citep[e.g.,][]{olson_2002, gubbins_2007, olson_2017, sahoo2020response} and low \citep{mound_2023} latitudes, and persistent offset of the axial dipole from the geographic pole \citep[e.g.,][]{olson_2012}; however, the specific locations of all of these features are seen to vary with the simulation control parameters. The presence of longitudinal structure in field morphology and secular variation has been found to extend deep into geological time, reflecting the long-lived nature of mantle heterogeneity \citep{biggin_2026}. Thermal CMB heterogeneity can also influence the frequency of polarity reversals \citep[e.g.,][]{olson_2010, frasson_2024, terra2024regionally}, and has sometimes been found to affect the relative secular variation (SV) in the Pacific and Atlantic hemispheres \citep{christensen_2003} and the properties of virtual geomagnetic pole paths \citep{kutzner2004simulated}, although this is not always the case \citep{korte_2022, mound_2023}. These studies considered geodynamo simulations driven by thermal buoyancy or a generalised ``codensity'' that combines the effects of heat and light element release due to solidification of the inner core. Although these approaches are theoretically and computationally convenient, they cannot account for the effects arising from the different diffusivities and boundary conditions of the thermal and chemical fields. The purpose of this study is to examine how these effects may influence the behaviour of the CMB magnetic field and the resultant signatures that could be observed at Earth's surface. 

Various regimes of two-component convection can arise, depending on the relative size of the diffusion coefficients and the component (temperature or chemical composition) that is stabilising \citep[e.g.,][]{turner1973buoyancy}. The bulk of Earth's outer core is expected to be unstable to both thermal and chemical convection \citep{davies_2011b, labrosse_2015}; however, the situation is less clear in the top $\mathcal{O}(100)$ km of the core. Uncertainties in the CMB heat flow and core thermal conductivity \citep[e.g.,][]{davies_2015, nimmo_2015,davies_2023} mean that thermally stabilising or destabilising conditions could prevail in the recent past (after formation of the inner core). The chemical flux is even more uncertain; some light elements are expected to dissolve into the core from the mantle, creating stable conditions \citep{buffett_2010,pozzo_2019}, while others may precipitate out, creating destabilising conditions \citep{orourke_2016,wilson_2022}. Here, we choose to focus on the ``top-heavy'' regime where both temperature and composition are destabilising throughout the core, but have different diffusivities and boundary conditions. Few previous studies have considered two-component convection simulations of the geodynamo \citep{tassin_2021,fan_2025,pruzina_2025}, and none have imposed laterally varying outer boundary conditions on the thermal field.

Here, we present 14 new thermochemical geodynamo simulations and investigate how lower‑mantle thermal heterogeneity and the balance of thermal versus chemical forcing may influence Earth's magnetic field. Model parameters are guided by our previous non-magnetic convection study \citep{naskar_2026} to target the rapidly-rotating turbulent flow regime, with varying thermal-to-chemical buoyancy forcing.
We model latent heat and light element release at the inner core boundary (ICB), with the thermal field assumed to diffuse $10$ times faster than the chemical field; although it may be significantly faster, $\mathcal{O}(10^2\text{--}10^3)$, in the core \citep{pozzo_2013}. We employ a tomographic heat flow pattern \citep{masters_1996} and thermal buoyancy profile used in our previous studies \citep{mound_2017, mound_2019, mound_2023, naskar_2026}. This thermal boundary condition may lead to the formation of regions of thermally stratified fluid, denoted as regional inversion lenses (RILs), at the top of the core beneath the African and Pacific Large Low Velocity Provinces (LLVPs) \citep{mound_2019,mound_2020}. We set the chemical field to be destabilising in the core (except for being neutral at the CMB), which might allow chemically-driven convection to disrupt the thermal RILs. On the other hand, previous studies have found that stratification due to light-element accumulation (LEA) can arise naturally from the internal dynamics of chemically dominated convection \citep{bouffard_2019} and, for the parameters we consider, LEA is expected to occur predominantly in polar regions \citep{naskar_2026}. Both low-latitude RILs and polar LEA might induce departures from geocentric axial dipole (GAD) behaviour, providing a potential observational constraint on the dynamic state of Earth's core. Therefore, we analyse the spatial and temporal variability of our geodynamo simulations across a range of timescales to elucidate the geomagnetic and paleomagnetic observable signatures of CMB thermal heterogeneity and relative thermal vs chemical forcing. The numerical model and simulation parameters are described in section \ref{sec:methods}, the simulation results are presented and compared to observational models in section \ref{sec:results},  discussed further in \ref{sec:discussion}, and the key findings are summarised in section \ref{sec:conclusion}.


\section{Methods}\label{sec:methods}

\subsection{Governing Equations}\label{sec:govequ}
We employ a numerical geodynamo model of a Boussinesq fluid \citep{willis_2007, davies_2011, matsui_2016}. A spherical coordinate system $(\radius,\theta,\phi)$ is used to represent the domain bounded by the inner and outer boundaries, with radii $\innerrad$ and $\outerrad$, respectively, with the shell gap $\shellthick = \outerrad - \innerrad$. The whole system rotates with a constant angular velocity $\rotationvec=\rotation\zhat$ about the vertical $\zhat$ axis, and gravity varies linearly with radius as $\grav= -\refgrav (\radius/\outerrad)\rhat$ where $\refgrav$ is the gravitational acceleration at the outer boundary. The fluid is a mixture of light components dissolved in a comparatively heavy liquid (e.g., oxygen mixed in liquid iron in Earth's outer core). The relevant physical properties of the mixture are the kinematic viscosity, $\kinvisc$, the thermal and chemical diffusivities, $\thermdiff$ and $\chemdiff$, and the coefficients of thermal and chemical expansion, $\thermexpan$ and $\chemexpan$. The thermal diffusivity is defined as $\thermdiff=\thermcond/\refdensity \specificheat$, where $\thermcond$ is the thermal conductivity, $\refdensity$ is the reference density, and $c_p$ is the specific heat of the mixture. The fluid is electrically conducting, with magnetic permeability $\refmagperm$ and magnetic diffusivity $\magdiff$.

The boundaries are assumed to be no-slip and electrically insulating at $\innerrad$ and $\outerrad$. Fixed-flux thermal boundary conditions are imposed at the inner and outer boundaries such that the total radial heat flow is equal through the inner and outer surfaces ($\boldsymbol{Q}_{T,i}=\boldsymbol{Q}_{T,o}$). The temperature gradient at the boundaries is expressed as $\boldsymbol{\nabla}T_{c}=-(\beta_{T}/\radius^{2})\rhat$ and related to the heat flux at the ICB through Fourier's Law as, $\boldsymbol{Q}_{T,i}=4\pi \innerrad^{2}(-\thermcond\boldsymbol{\nabla}T_{c})=4\pi \thermcond \beta_T \rhat$. A standard setup \citep[e.g.,][]{kutzner_2002} is used to model the release of light elements from the inner core boundary, with fixed flux conditions imposed at $\innerrad$ such that $\boldsymbol{\nabla}\xi_{c,i}=-(\beta_{\xi}/r_i^{2})\rhat$. To ensure stationary solutions, we assume that the chemical flux from the inner core is balanced by a spatially homogeneous sink ($\chemsource$) that maintains the global balance of lighter elements \citep[see e.g.,][]{kono_2001}.
 
The nondimensional governing equations determining the velocity field ($\vel$), the magnetic field ($\magfield$), temperature ($T$) and chemical composition ($\xi$) can be found in equations \ref{eqn:continuity_nd}-\ref{eqn:composition_nd} of the Supplementary Information. The non-dimensional numbers appearing in these equations are the Ekman number ($E$), the thermal and chemical flux Rayleigh numbers ($\RaT$ and $\RaC$), and the thermal, chemical, and magnetic Prandtl numbers ($\PrT$, $\PrC$ and $Pm$) defined as
\begin{equation}\label{eqn:nd_parameters}
\begin{split}
\Ek = \frac{\kinvisc}{2\rotation \shellthick^2}\qquad
\RaT = \frac{\refgrav\thermexpan\beta_T \shellthick^3}{\kinvisc \thermdiff \outerrad} \qquad
\RaC = \frac{\refgrav\chemexpan\beta_\xi \shellthick^3}{\kinvisc \chemdiff \outerrad} \\
\PrT = \frac{\kinvisc}{\thermdiff} \quad
\PrC = \frac{\kinvisc}{\chemdiff} \quad
\Pm = \frac{\kinvisc}{\magdiff}
\end{split}
\end{equation}
The modified flux Rayleigh number used in the subsequent sections relates to the flux Rayleigh number as $\tRaT=\RaT\Ek$ (and similarly $\tRaC=\RaC\Ek$). To isolate the effect of CMB heat flux heterogeneity, we run simulation pairs with homogeneous thermal boundary conditions and laterally heterogeneous thermal flux at the CMB, following \citet{mound_2017}. The parameter space now also includes the pattern and amplitude of lateral variation in the CMB heat flux.  In this study, the pattern is fixed using an inferred heat flux pattern from seismic tomography following \citet{masters_1996}. The amplitude of heterogeneity is characterised as $\qstar=(q^{T}_{max}-q^{T}_{min})/q^{T}_{avg}$   where $q^{T}_{max}$,  $q^{T}_{min}$, and $q^{T}_{avg}$ are the maximum, minimum, and horizontally averaged heat flux through the CMB, respectively. The amplitude is also fixed at $\qstar= 5$, where $\qstar>2$ has been reported to produce RILs for pure thermal convection \citep{mound_2019}. The homogeneous models correspond to $\qstar=0$. The average heat flux $q^{T}_{avg}$ is the same for homogeneous and heterogeneous model pairs.

\subsection{Numerical details}
We have performed $14$ new geodynamo simulations by varying $\tRaT$, $\tRaC$ and $\qstar$ while keeping the other parameters fixed at  $\innerrad/\outerrad=0.35$, $\Ek=10^{-5}$, $\PrT = 1$, $\PrC = 10$ and $Pm = 5$. The aspect ratio $\innerrad/\outerrad=0.35$ is representative of Earth’s present-day core configuration. The Ekman number is chosen small enough to ensure rotationally constrained dynamics, while allowing sufficient runtime to assess the typical magnetic field behaviour.  We also adopt the common practice of keeping $\PrT = 1$, although this is considerably higher than the estimated value for the core ($\PrT \sim 0.05$, see \citet{pozzo_2013}). Conversely, our choice of $\PrC = 10$ is considerably lower than the core estimate ($\PrC \sim 100$,\ \citet{pozzo_2013}), and is selected solely for computational benefit. The Prandtl number ratio $\PrC/\PrT$, which controls the scale separation between the two scalar fields, is therefore much lower in our simulations compared to Earth's core.    

Of the $14$ geodynamo simulations, $7$ are run with homogeneous CMB heat flux conditions ($\qstar = 0$). Four of these simulations with $\tRaT \in \{90, 550, 1200, 4000\}$ are run at $\tRaC=10^{4}$ and three with $\tRaT \in \{90, 550, 1200\}$ are run at $\tRaC=10^{5}$. At $\Ek=10^{-5}$, onset of pure thermal convection with $\PrT=1$ and pure chemical convection with $\PrC=10$ occurs at $\cRaT = 24.7$ and $\cRaC = 34.1$, respectively \citep{naskar_2025b}. The range of Rayleigh numbers are chosen from our previous parameter survey of non-magnetic double-diffusive convection simulations at $\Ek=10^{-5}$ \citep{naskar_2026}, which identifies the dynamical regime where vigorous turbulence coexists with strong rotational constraint. Along with $7$ homogeneous cases, $7$ more simulations are run for the same ($\tRaT$, $\tRaC$) pairs, with a tomographic pattern of heat flux at the CMB with $\qstar = 5$. In the subsequent sections, the simulations with homogeneous CMB heat flux conditions (i.e. $\qstar=0$), and those with a tomographic pattern of CMB heat flux with $\qstar = 5$ are simply referred to as homogeneous and heterogeneous simulations, respectively. The simulation parameters are listed in \ref{app:diagnostics}. 

Hyperdiffusion has been used to enable long simulation runs \citep[see e.g.,][]{aubert_2017}, where enhanced diffusivity is applied to scales smaller than a certain "cut-off" length-scale, represented by a spherical harmonic degree $\ell_h=128$. The diffusion operator applied to the momentum, composition, and magnetic field remains as $\nabla^2$ for $\ell<\ell_h$, and is replaced by $q_h^{\ell-\ell_h}\nabla^2$ for $\ell\geq\ell_h$. The value of the parameter $q_h=1.0325$ is chosen to ensure a smooth increase of hyperdiffusion with wavenumber. After some initial transient, all simulations have been run for at least $0.2$ magnetic diffusion times, which has been reported to be long enough to determine the signature of CMB heterogeneity in the time-average morphology of the magnetic field \citep{davies_2014, mound_2015}. 


\subsection{Global diagnostics}\label{sec:diagnostics}
 We use the following dimensionless output diagnostics to analyse the simulations. The global root mean square flow velocity is represented by the magnetic Reynolds number
 \begin{equation}
   \Rm = \frac{\charvel \shellthick}{\magdiff} = \sqrt{\frac{2 \kinEnergy}{\volshell}},
\end{equation}
where $\charvel$ is the characteristic velocity, $\volshell$ is the volume of the outer core, and the kinetic energy is defined as

\begin{equation}
    \kinEnergy = \frac{1}{2} \int \vel^2 dV .
\end{equation}
The magnetic Reynolds number is related to the Reynolds number through $Re=\charvel \shellthick/\kinvisc=Rm/Pm$. Also, the total buoyant power ($P_{tot}$) is defined as,

\begin{equation}
    \totbuopow = \int_{\volshell} \Pm\left(\frac{\widetilde{Ra}_{T}}{Pr_T}\tT+\frac{\widetilde{Ra}_{\xi}}{Pr_\xi}\tC\right)ru_{r} dV ,
    \label{eq:totbuopow}
\end{equation}
where $u_r$ is the radial velocity component. Additionally, the Elsasser number $\elsasser$ measures the dimensionless magnetic energy $\magEnergy$ and is given by

\begin{equation}
    \elsasser = \frac{\charmag^2}{2\rotation\refdensity\refmagperm\magdiff} = \frac{2\magEnergy \Ek}{\Pm \volshell} ,
    \label{eq:elsasser}
\end{equation}
where
\begin{equation}
    \magEnergy = \frac{\Pm}{2\Ek} \int B^2 dV .
\end{equation}

Finally, dipolarity of the magnetic field at the outer boundary is calculated  by 
\begin{equation}
 f_\text{dip} = \left(
    \frac{\int \magfield_{\ell=1}(\radius=\outerrad)\cdot\magfield_{\ell=1}(\radius=\outerrad) dA}
    {\int \magfield_{\ell\leq 12}(\radius=\outerrad)\cdot\magfield_{\ell\leq 12}(\radius=\outerrad) dA} 
    \right)^{1/2}.
\end{equation}

\subsection{Force diagnostics}\label{sec:force}
Earth's core is expected to exhibit a primary quasi-geostrophic balance between Coriolis and Pressure forces, with the residual Coriolis force balanced by Lorentz and buoyancy forces in the secondary balance, constituting a QG-MAC balance of forces governing the dynamics, while inertial and viscous effects remain small \citep{aurnou_2017, calkins_2018, davidson_2013,yadav_2016,aubert_2017,aubert_2019,schwaiger_2019,schwaiger_2021}. We estimate the force balance in our simulations from the scale-dependent forces along with the volume-averaged ratio of various force terms in the fluctuating momentum equation (equation \ref{eqn:momentum_f}), following \citet{naskar_2025b}. In \citet{naskar_2025b}, the forces and curled forces are partitioned into mean (i.e. azimuthal average) and corresponding fluctuating parts (see also e.g.\citet[][]{calkins_2021,nicoski_2024}), and we only consider the fluctuating part of the forces here (indicated by a prime). All the forces are integrated over the bulk fluid, excluding regions within a radial thickness corresponding to $10$ times the Ekman-layer depths adjacent to the upper and lower boundaries (defined using the linear intersection method). 
The metrics to verify the dynamical balances are based on the ratio of the magnitudes of the various forces, and their curls \citep{naskar_2025b,clarke_2026}. The inertial contribution to the primary balance is quantified by the ratio of inertia to Coriolis forces $\FIC$, and the ratios of curled Lorentz (and Archimedean) to Coriolis forces $\CFMC$ (and $\CFAC$) have been used to check the secondary balance. Here, the curled force representation ensures reliable estimates of the secondary forces by eliminating the dynamically irrelevant gradient parts of the forces that are balanced by the pressure gradient\citep{teed_2023,teed_2025}. The metrics are defined as

\begin{equation}\label{eqn:FIC}
\begin{split}
\FIC = \frac{|\boldsymbol{I'}|}{|\boldsymbol{C'}|}, ~~~
\CFMC = \frac{|\boldsymbol{\nabla\times M'}|}{|\boldsymbol{\nabla\times C'}|}, ~~~
\CFAC = \frac{|\boldsymbol{\nabla\times A'}|}{|\boldsymbol{\nabla\times C'}|},
\end{split}
\end{equation}
where the total buoyancy is $\boldsymbol{A'}= \boldsymbol{A'_T} + \boldsymbol{A'_\xi}$. Additionally, we use another metric to estimate the relative importance of two buoyancy forces,

\begin{equation}\label{eqn:FACT}
\FACT = \frac{|\boldsymbol{A'_T}|}{|\boldsymbol{A'_\xi}|}.      
\end{equation}

\subsection{Measures of field morphology and variability}\label{sec:compliance}
We compute existing criteria to quantify the spatial structure and temporal variability of the simulated magnetic fields and compare them to geomagnetic and paleomagnetic observations across various time scales. Our aim is to assess differences between homogeneous and heterogeneous simulations, and between thermally dominated and chemically dominated simulations. We further seek to evaluate whether such differences are reflected in observed features of Earth's magnetic field.
We consider the gufm1 field model \citep{jackson_2000} spanning the past 400 years, the CALS10k.2 model \citep{constable_2016} of the last 10,000 years, and the GGF100k model \citep{panovska_2018} spanning the last 100,000 years. This range of timescales is both numerically achievable with the present set of simulation parameters and sufficient to investigate the signature of CMB heterogeneity in the non-dipole time-averaged field (TAF) \citep{davies_2014}.  To compare with the field models, the magnetic diffusive timescale in the simulations are rescaled assuming a magnetic diffusivity of $\eta = 1.6\ \text{m}^2/\text{s}$, which corresponds to a magnetic diffusion time of roughly 100,000 years.
We compare simulated and observed fields based on:
 \begin{itemize}
     \item Temporal field variability on decadal timescales using the master secular variation time-scale, $\tau_{SV}$, following \citet{lhuillier_2011}; 
     \item Morphological semblance on the timescale of centuries using the four compliance criteria proposed by \citet{christensen_2010}; 
     \item $10-100$~kyr variability using the paleo-secular variation index ($P_i$),  with $\truncation=10$ \citep{panovska_2017};
     \item Statistical properties of field morphology and temporal variability over $>100$~kyr timescales using the five $\QPM$ criteria proposed by \citet{sprain_2019}.      
 \end{itemize}
Precise definitions of these criteria are given in section \ref{sec:compliance_SI} of the SI, though they are identical to those used in the original publications. 

Thermal RILs that form under LLVPs or light elements that accumulate preferentially near the poles may imprint regional signatures on the magnetic field. If sufficiently strong and persistent, such signatures would be expressed as departures from a geocentric axial dipole field in the long-term behaviour of the Earth. Therefore, we look for measures of such behaviour that are both potentially observable and quantitatively distinguishable between homogeneous and heterogeneous (or thermally/chemically dominated) cases. 
In particular, we consider the following:

\begin{itemize}
    \item The dipole-dominance and longitudinal variation of the TAF at Earth's surface quantified using the ratios $\ADNADTAF$ and $\ZNDNDTAF$, calculated from the Lowes power spectra for the magnetic energy \citep{lowes_1974}, following \citet{biggin_2026};
    
    \item The polar minima in the TAF quantified by the time-averaged radial field magnitude at the normalised by its maximum value in the northern hemisphere; 
    
    \item Deviation from GAD measured by the ratios of specific Gauss coefficients relative to the axial dipole (e.g., $\GC{1}{1}$ and $\GC{3}{0}$).
    
    \item Eccentricity of the dipole, calculated as an offset from the centre of Earth in the equatorial plane ($\dipecc$) \citep{james_1967,olson_2012}.
    
    \item  Longitudinal structures near the equator \citep{mound_2023} and their variability, quantified using the median ($\Smed$) and inter-quartile range ($\Siqr$) of dispersion of virtual geomagnetic poles (VGPs) at low latitudes following \citet{biggin_2026}.
    
    \item Longitudinal variation of the time- and latitudinally-averaged paleo-secular variation index $P_i$ following \citep{mason_2024}.
    
    \item The magnitude and spatial variability of the inclination anomaly ($\IncAnom$), on Earth's surface. The longitudinal variability of inclination anomaly is calculated from the time-averaged $\IncAnom$ map at the surface. At each latitude, we define the range of $\IncAnom$ as the difference between the maximum and minimum $\IncAnom$ in the longitudinal distribution.
\end{itemize}

All of the metrics outlined in this section will not be discussed in detail below; however, they are reported in full in \ref{app:diagnostics} to assist further comparison with previous work.

\section{Results}\label{sec:results}

\subsection{Global diagnostics}\label{sec:results_diag}
\begin{figure}[h!]
\centering
\begin{overpic}[width=0.8\linewidth,trim={3cm 9cm 2cm 9.8cm},clip]{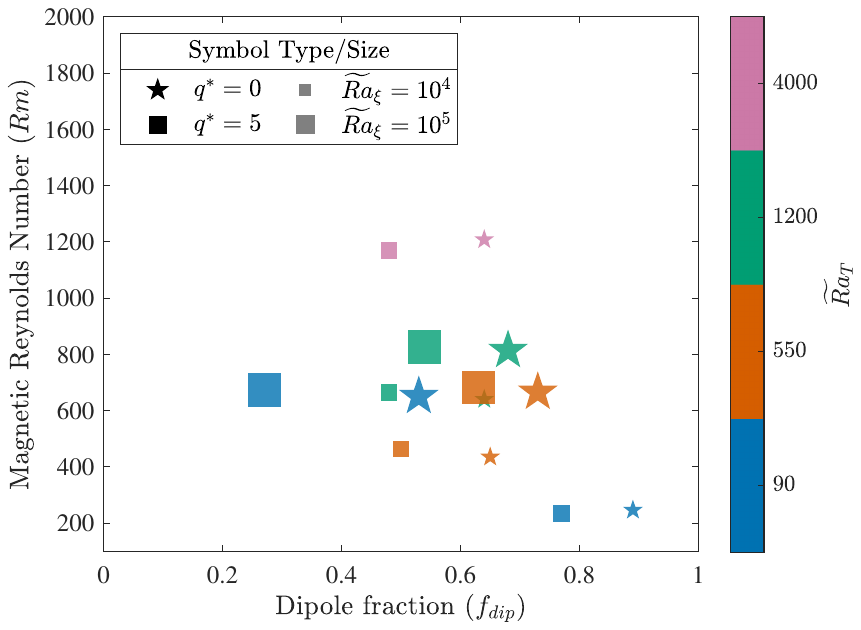}
\end{overpic}
\caption{All 14 simulations in the $\Rm-\fdip$ plane, coloured by thermal Rayleigh number ($\tRaT$). Symbol shape demarcates homogeneous (pentagram) and heterogeneous (square) simulations, whereas symbol size differentiates between low (small) and high (big) $\tRaC$ cases.}\label{fig:f_dip}
\end{figure}

 We begin by examining several global diagnostic quantities to characterise the dynamo runs. Generally, the simulations produce dipole-dominated fields with dipole fractions $\fdip \geq 0.48$ (Figure~\ref{fig:f_dip}), except for one heterogeneous case ($\qstar = 5$, $\tRaT = 90$ and $\tRaC = 10^{5}$) with $\fdip=0.27$ that exhibits multipolar reversing behaviour. Most of our simulations access Earth-like values of the magnetic Reynolds number $\Rm$, which is estimated to vary between $\sim 500$ and several thousand in Earth's core depending on assumed values of the core's electrical conductivity and the root mean square flow speed \citep{davies_2015}.   
 In our suite, the heterogeneous simulations (square symbols, Figure~\ref{fig:f_dip}) systematically yield smaller $\fdip$ values compared to their homogeneous counterpart, despite having comparable $Rm$.
 The magnetic-to-kinetic energy ratio, $\Mratio=\magEnergy/\kinEnergy$, which is large in Earth's core, ranges between $2.9\text{---}39.4$ in our simulations (see \ref{app:diagnostics}) with no clear trend as the control parameters are varied.
 With the exception of three simulations at $\tRaT = 90$, the suite of simulations sits within a range of $\fdip$ and $\Rm$, that has been found to produce magnetic fields with good compliance with the observed characteristics of Earth's magnetic field \citep{nakagawa_2022}. 
 
\subsection{Force balance}\label{sec:results_force}
The dynamical regime of our simulations can be further assessed using scale-dependent force spectra as well as volume-averaged force ratios \citep{naskar_2025b}. In the heterogeneous simulations (e.g., Figure~\ref{fig:fspec}b), thermal buoyancy is systematically enhanced at spherical harmonic degree $\ell=2$, the dominant scale of the imposed thermal heterogeneity, relative to their homogeneous counterparts (e.g., Figure~\ref{fig:fspec}a). Furthermore, simulations with high $\tRaC$ (e.g., Figure~\ref{fig:fspec}c) show stronger chemical buoyancy with the crossover between dominant buoyancy and Lorentz force occurring at a smaller spatial scale (i.e. higher $\ell$) than corresponding low $\tRaC$ cases (e.g., Figure~\ref{fig:fspec}b). Nevertheless, across all cases, the simulations exhibit the characteristic QG-MAC balance \citep{aubert_2017,schwaiger_2019, nakagawa_2022}, regardless of the heat flux boundary condition or the dominant buoyancy source.

The dependence of the global force balance on the control parameters can be quantified with volume-averaged force measures \citep{naskar_2025b,teed_2025,clarke_2026}. The integrated ratios of curled Lorentz to curled Coriolis forces remain near unity ($\CFMC\sim1$), indicating strong field solutions for all cases \citep[see also][]{teed_2025}. Furthermore, the curled total buoyancy to curled Coriolis forces ($\CFAC$) fall within $\mathcal{O}(0.1-1)$ across the parameter range examined (\ref{app:diagnostics}), consistent with a secondary MAC balance. The ratio of inertia to Coriolis forces ($\FIC$) increases with convective driving in the simulations but remains $\mathcal{O}(0.01-0.1)$, confirming that inertia plays a subdominant dynamical role. The ratio of thermal to chemical buoyancy force ($\FACT$) varies by nearly three orders of magnitude over the parameter space explored here (\ref{app:diagnostics}), spanning regimes from thermally dominated convection (e.g., $\FACT = 41.5$ at $\tRaT=4000$, $\tRaC=10^4$ and $\qstar=0$) to chemically dominated convection (e.g., $\FACT = 0.09$, at $\tRaT=90$, $\tRaC=10^5$ and $\qstar=0$).


\subsection{Magnetic field morphology and variability}\label{sec:results_compliance}


We now consider metrics describing the general spatial morphology and time variability of the simulated magnetic fields, with particular attention to whether systematic differences arise between homogeneous and heterogeneous simulations, or between thermally and chemically dominated regimes. These metrics were developed taking into account constraints from different observational models and thus provide complementary expectations for field behaviour over a range of timescales (recall Section \ref{sec:compliance}). 

The SV timescale, $\tausv$, applies to relatively short-term temporal variability and decreases with increasing thermal or chemical forcing in our simulations. The homogeneous simulations have $\tausv$ values that are somewhat (approximately $5-15$\%) smaller than their heterogeneous counterparts; however, all simulations follow an approximately inverse scaling relation between $\tausv$ and the magnetic Reynolds number (Figure \ref{fig:Rm_vs_tau_SV}). An empirical fit to the data gives coefficients closely matching the scaling predictions of \citet[][]{christensen_2012} (see their equation $18$). 


\begin{figure}[h!]
\centering
\begin{overpic}[width=0.49\linewidth,trim={0cm 0cm 0cm 0cm},clip]{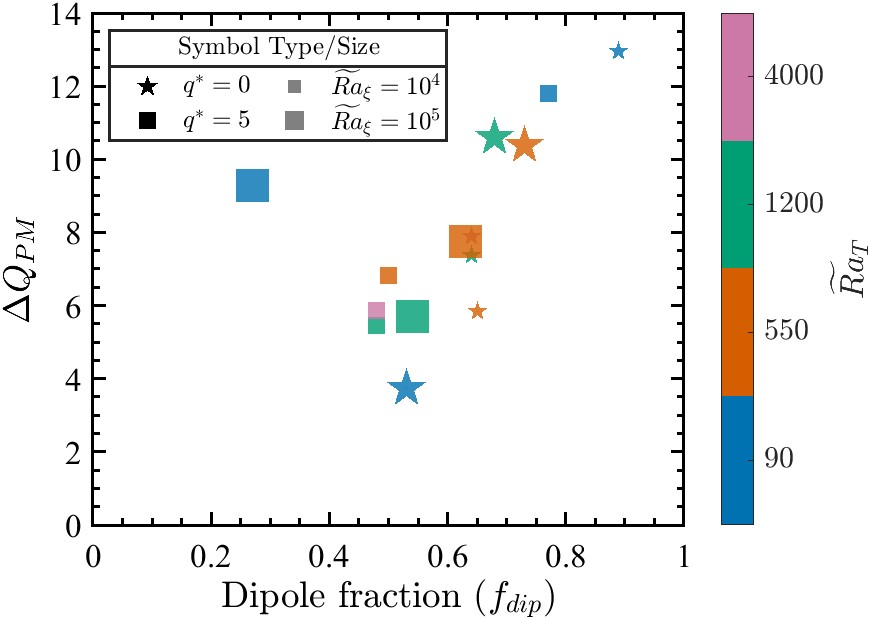}
\put(-2,73){$(a)$}
\end{overpic}
\begin{overpic}[width=0.47\linewidth,trim={0cm 0cm 0cm 0cm},clip]{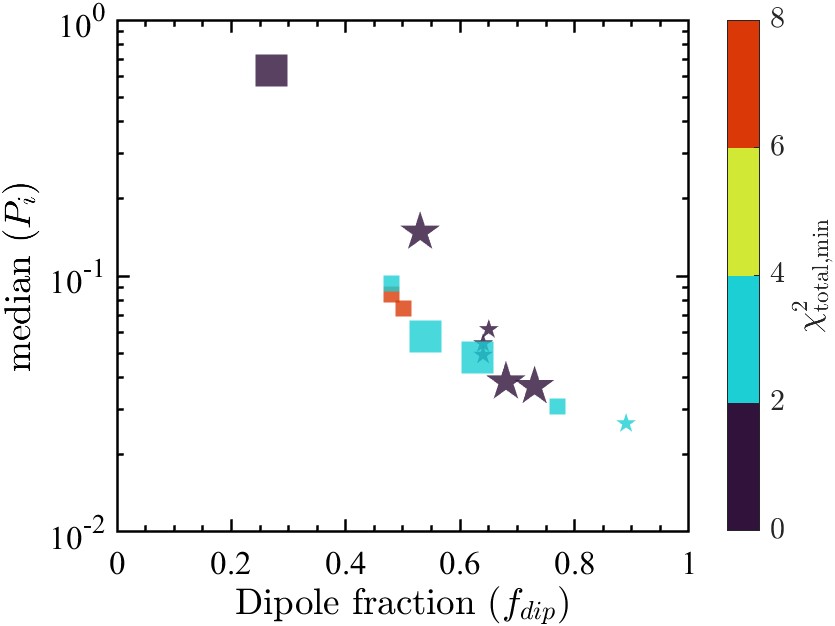}
\put(-2,75){$(b)$}
\end{overpic}
\caption{Variation of the total (a) $\QPM$ misfit ($\DelQPM$) and (b)  median paleo-secular variation index $P_i$ with dipole fraction ($\fdip$). Symbols in (a) have the same meaning as in Figure \ref{fig:f_dip}, while symbols in (b) are colored by the minimum value of total misfit  ($\chisqtotmin$) as calculated from morphological semblance criteria of \citet[][]{christensen_2010}.}\label{fig:compliance_vs_fdip}
\end{figure}

Comparison to the spatial morphology of Earth's recent magnetic field is covered by the $\chisqtot$ compliance computed following \citet{christensen_2010}. Time-averaged values for the homogeneous simulations generally have a lower average $\chisqtot$ than their heterogeneous counterparts (\ref{app:diagnostics}). This contrast is mainly due to the propensity of heterogeneous simulations to have patches of concentrated flux that are stronger than both the homogeneous simulations and those inferred for modern Earth. However, since the flux concentration factor (see section \ref{sec:compliance_SI}) is not well constrained for the geomagnetic field on millennial and longer timescales \citep{panovska_2019, wardinski_2025}, this does not necessarily imply a shortcoming of the heterogeneous simulations. Indeed, most (12 of 14) simulations have periods of time for which they are in good or excellent morphological agreement with the modern geomagnetic field (i.e., $\chisqtotmin<4$). Other than the formation of strong flux patches in heterogeneous simulations, the $\tausv$ and $\chisqtot$ metrics, derived based on models of Earth's modern geomagnetic field, indicate relatively small differences between homogeneous and heterogeneous simulations. Furthermore, the simulations can achieve excellent compliance ($\chisqtotmin \leq 2$) regardless of whether the driving mechanism is thermal or compositional, similar to the findings of \citet{tassin_2021}.

The long-term variability of the magnetic field can be characterised by the median of paleo-secular variation index \citep{panovska_2017}, which decreases with increasing dipolarity of the simulated fields (Figure~\ref{fig:compliance_vs_fdip}b). 
Conversely, the $\DelQPM$ misfits (Figure \ref{fig:compliance_vs_fdip}a) generally decrease with decreasing $\fdip$ reaching values of $\DelQPM \sim 6-8$, around $\fdip \sim 0.5-0.6$. The multipolar reversing case departs from this trend, exhibiting a much lower $\fdip$. A similar dependence of misfit with $\fdip$ was identified by \citet{meduri_2020}, who reported a minimum $\DelQPM \sim 5$ near $\fdip \sim 0.5-0.6$. However, given that our simulations are considerably shorter ($\sim20\text{---}100$ kyr) than the paleofield observations on which the $\QPM$ criteria are based, we don't place emphasis on absolute $\DelQPM$ values and instead focus on inter-simulation comparisons. Since $\fdip$ for the homogeneous simulations is systematically higher than their heterogeneous counterparts, the homogeneous simulations have systematically lower values for their median $P_i$ and systematically larger values of $\DelQPM$. No systematic dependence of these metrics on the relative thermal versus chemical forcing is evident.


\subsection{Regional magnetic signatures}\label{sec:results_TAF}
\begin{figure}[h!]
\centering
\begin{overpic}[width=0.23\linewidth,trim={0cm 0cm 0cm 1.5cm},clip]{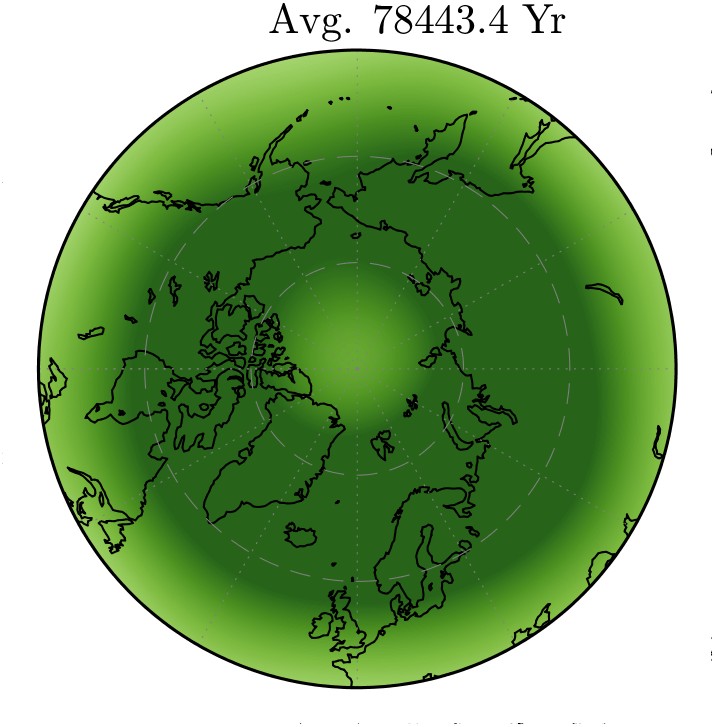}
\put(-2,100){$(a)\,\qstar=0,\ \tRaC=10^4$}
\end{overpic}
\begin{overpic}[width=0.48\linewidth,trim={0cm 0cm 0cm 0cm},clip]{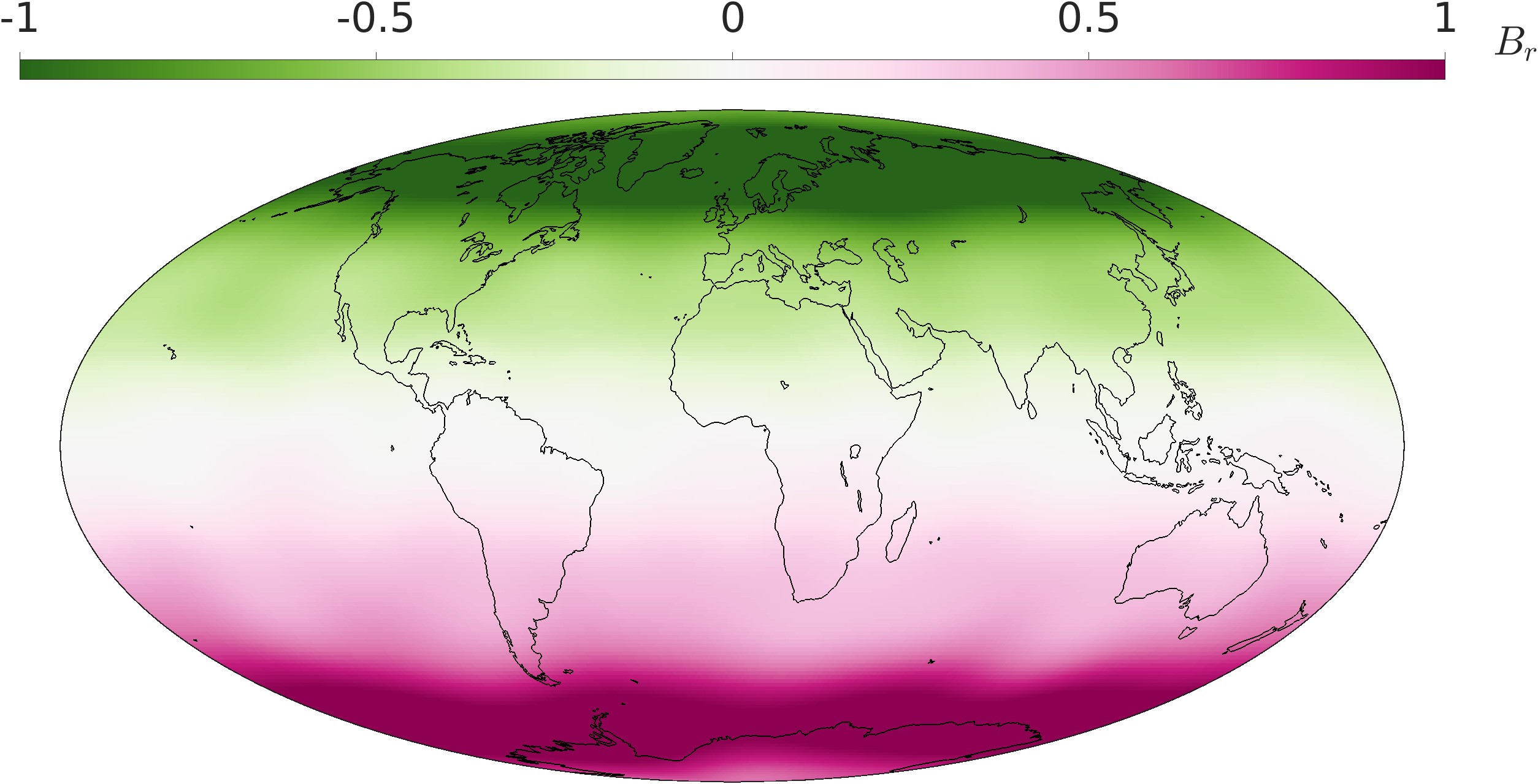}  
\end{overpic}
\begin{overpic}[width=0.23\linewidth,trim={0cm 0cm 0cm 1.5cm},clip]{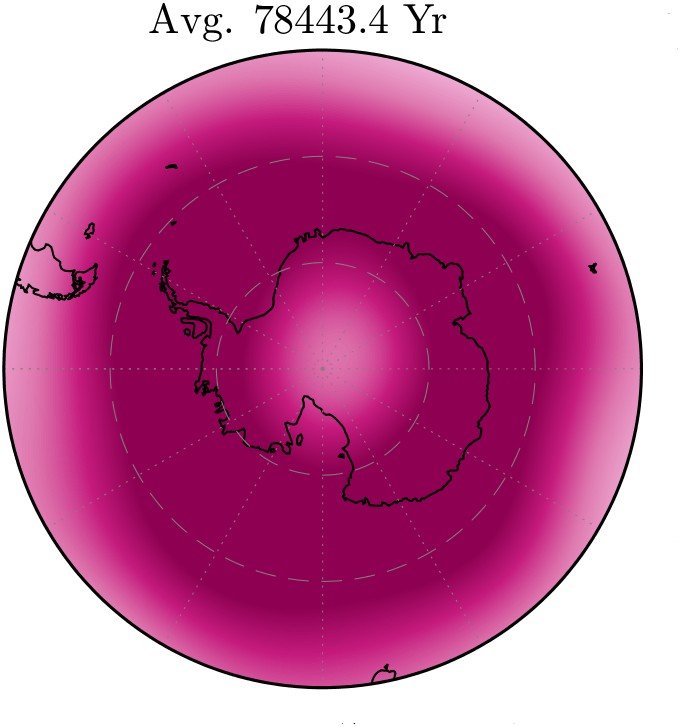}  
\end{overpic}

\vspace{5mm}

\begin{overpic}[width=0.23\linewidth,trim={0cm 0cm 0cm 1.5cm},clip]{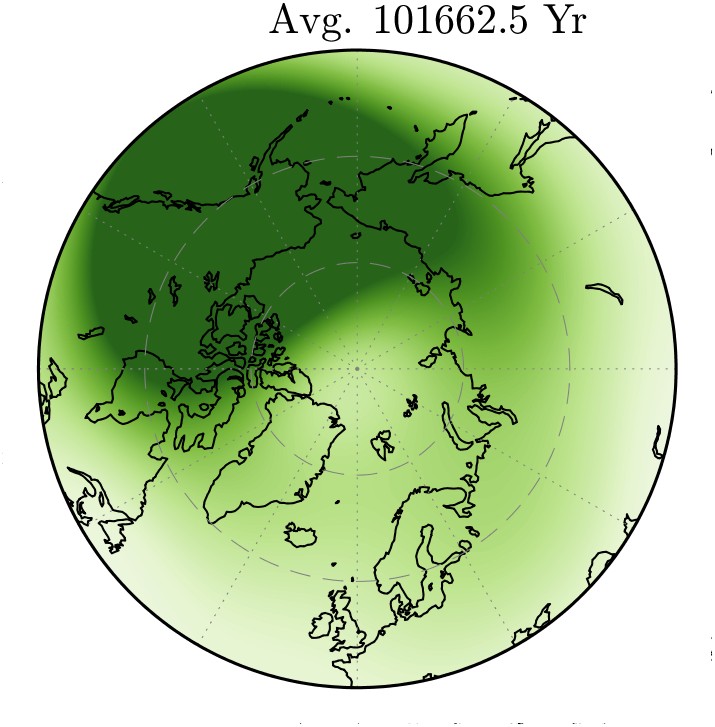}
\put(-2,100){$(b)\,\qstar=5,\ \tRaC=10^4$}
\end{overpic}
\begin{overpic}[width=0.48\linewidth,trim={0cm 0cm 0cm 5cm},clip]{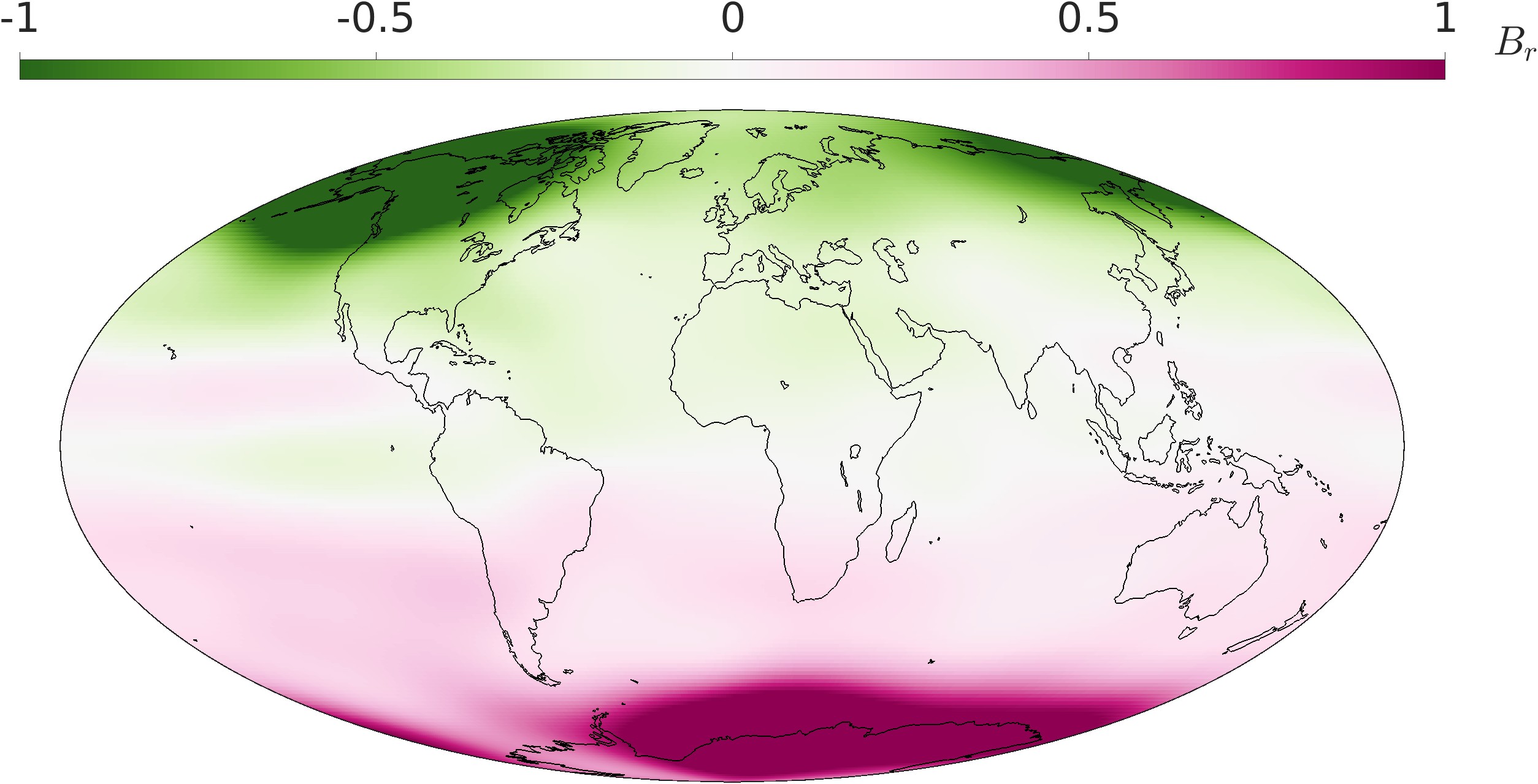}  
\end{overpic}
\begin{overpic}[width=0.23\linewidth,trim={0cm 0cm 0cm 1.5cm},clip]{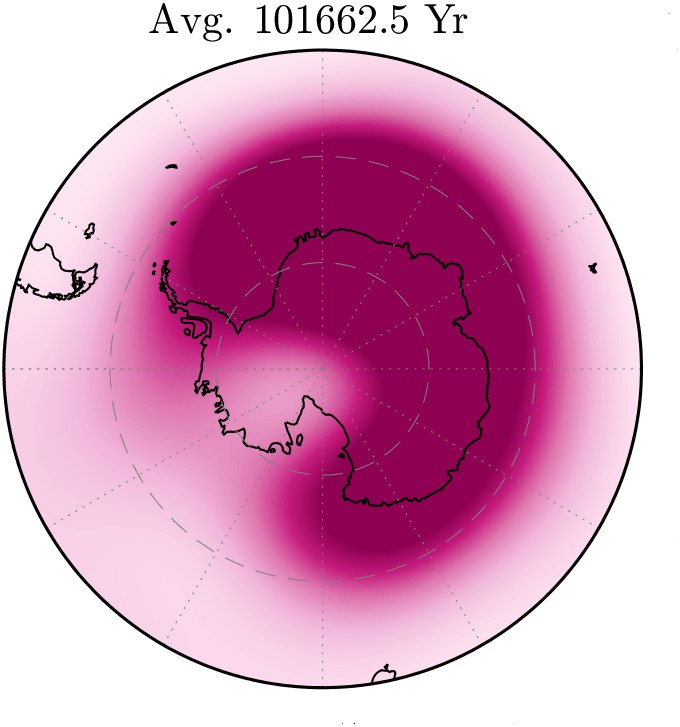}
\end{overpic}

\vspace{5mm}

\begin{overpic}[width=0.23\linewidth,trim={0cm 0cm 0cm 1.5cm},clip]{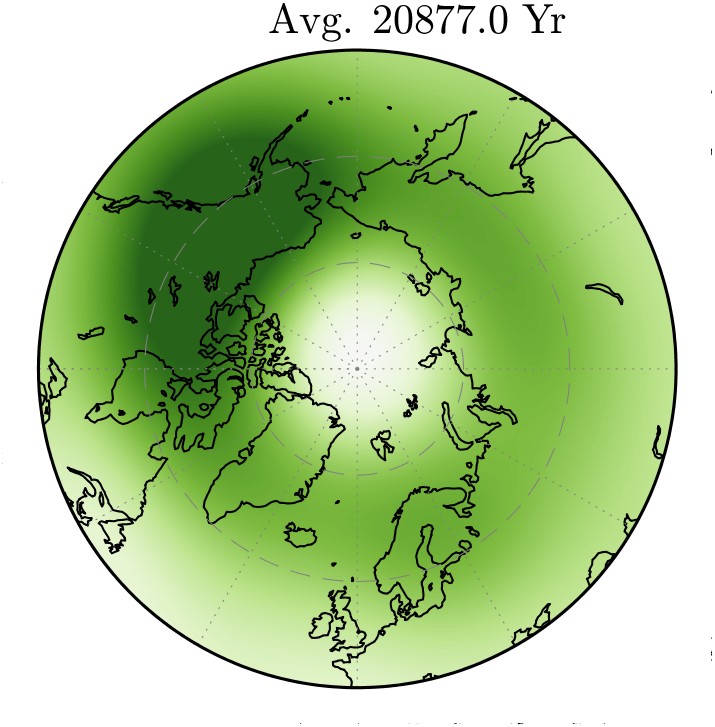}
\put(-2,100){$(c)\,\qstar=5,\ \tRaC=10^5$}
\end{overpic}
\begin{overpic}[width=0.48\linewidth,trim={0cm 0cm 0cm 5cm},clip]{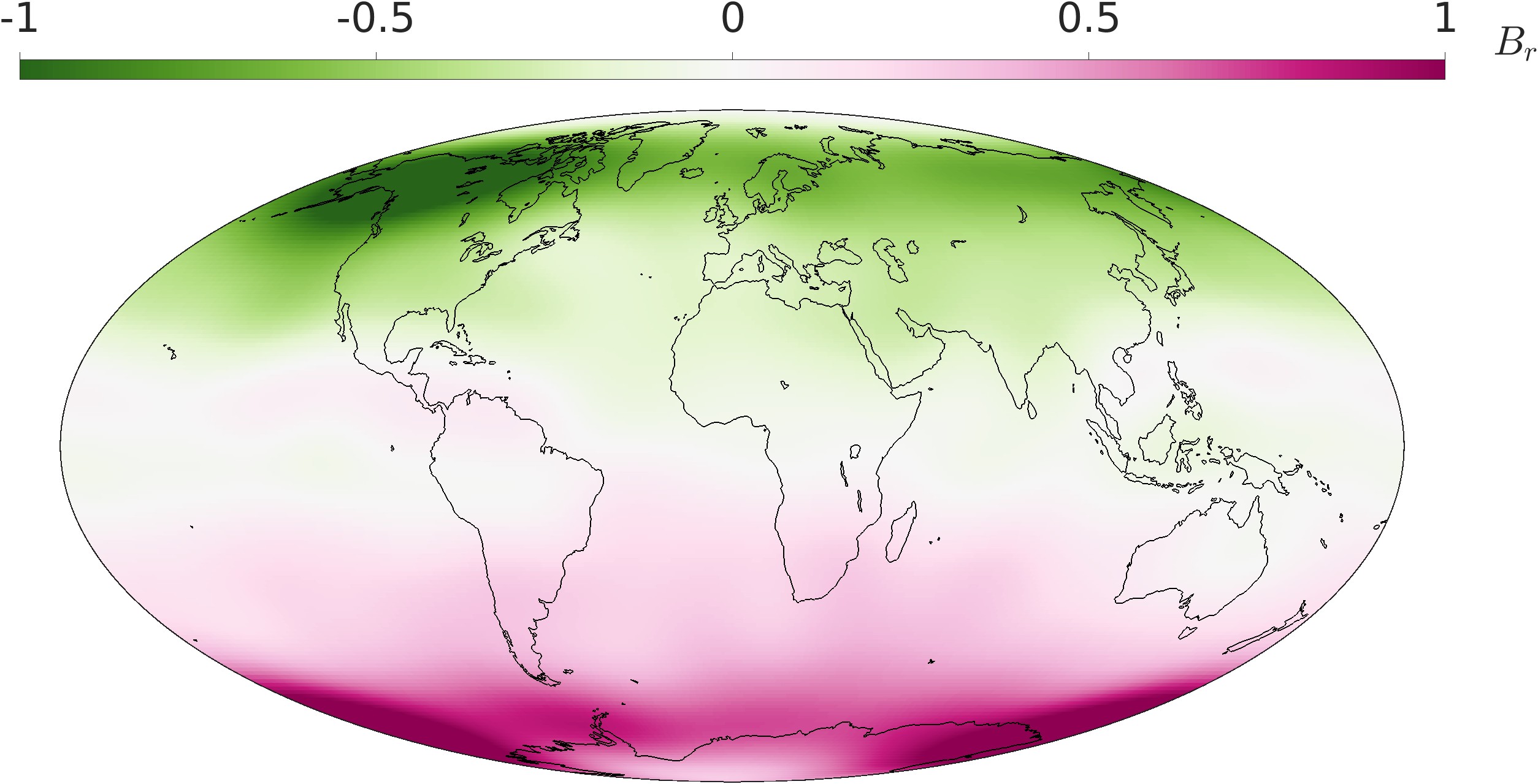}  
\end{overpic}
\begin{overpic}[width=0.23\linewidth,trim={0cm 0cm 0cm 1.5cm},clip]{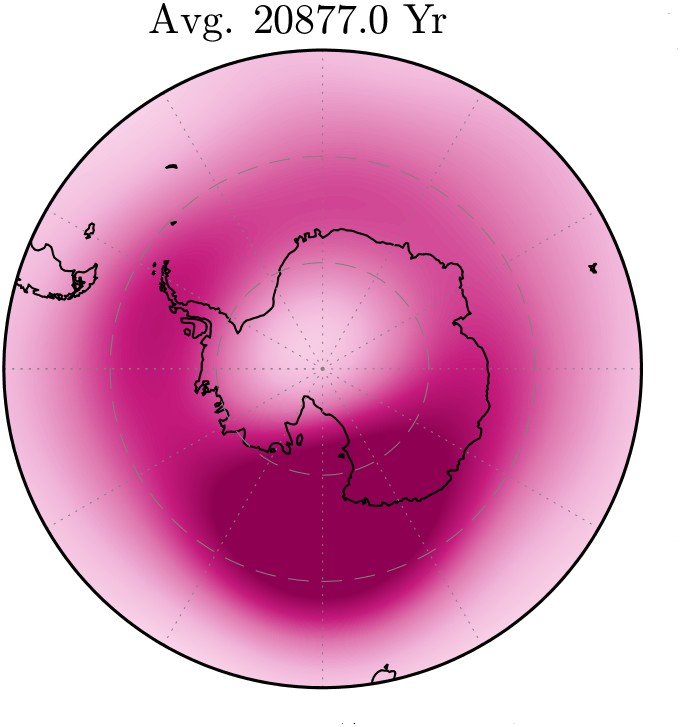}  
\end{overpic}
\caption{Time-averaged radial magnetic field at the CMB truncated at $\truncation=12$ for three simulations at $\tRaT=550$ with (a) $\qstar=0$, $\tRaC=10^{4}$ (b) $\qstar=5$, $\tRaC=10^{4}$ and (c) $\qstar=5$, $\tRaC=10^{5}$. Left (north) and right (south) panels show polar views down to $45^\circ$ latitude.}\label{fig:TAF_map}
\end{figure}

The criteria so far considered are based on global measures of the magnetic field's spatial morphology or temporal variability, and thus are not well suited to detect the potential regional signatures of CMB heat flux heterogeneity or preferential LEA in polar regions.
To illustrate the typical differences between our simulations, we focus on three simulations: one homogeneous ($\qstar=0,\ \tRaC=10^4$) and two heterogeneous cases ($\qstar=5,\ \tRaC=10^4$ and $\tRaC=10^5$), all at fixed $\tRaT = 550$. The time-averaged field of the $\tRaC=10^{4}$ heterogeneous model (Figure~\ref{fig:TAF_map}b) exhibits significant longitudinal variation compared to the corresponding homogeneous model (Figure~\ref{fig:TAF_map}a). This propensity for heterogeneous boundary forcing to induce longitudinal field structure holds across all our simulations and agrees with previous results for purely thermal simulations \citep{olson_2002,olson_2018,mound_2023,biggin_2026}. 

The most prominent signature of the heterogeneous simulations is the presence of high-latitude magnetic flux patch(es) in the TAF that are absent in the homogeneous simulations; although the heterogeneous simulations also exhibit longitudinal structures near the equator, as found by  \citet{mound_2023}. The precise structure and position of the longitudinal features in the TAF vary with the simulation control parameters; however, their importance can be estimated by the ratio of axial dipole to non-axial dipole power in the time-averaged field, $\ADNADTAF$. If we discount the case with $\fdip = 0.27$, the homogeneous simulations have consistently larger values of this ratio than their heterogeneous counterparts (by a factor of $\sim 5 - 20$). 
Additionally, all of our simulations exhibit a polar minimum in the radial magnetic field ($B_r$), which tends to deepen with increasing $\tRaC$ and simulations dominated by chemical buoyancy tend to have a more pronounced polar minimum than thermally dominated simulations at comparable total buoyant power (Figure~\ref{fig:pol_min}). Also, heterogeneous simulations have a stronger polar minimum than their homogeneous counterparts.

\begin{figure}[h!]
\centering
\begin{overpic}[width=0.7\linewidth,trim={0cm 8cm 3.8cm 7cm},clip]{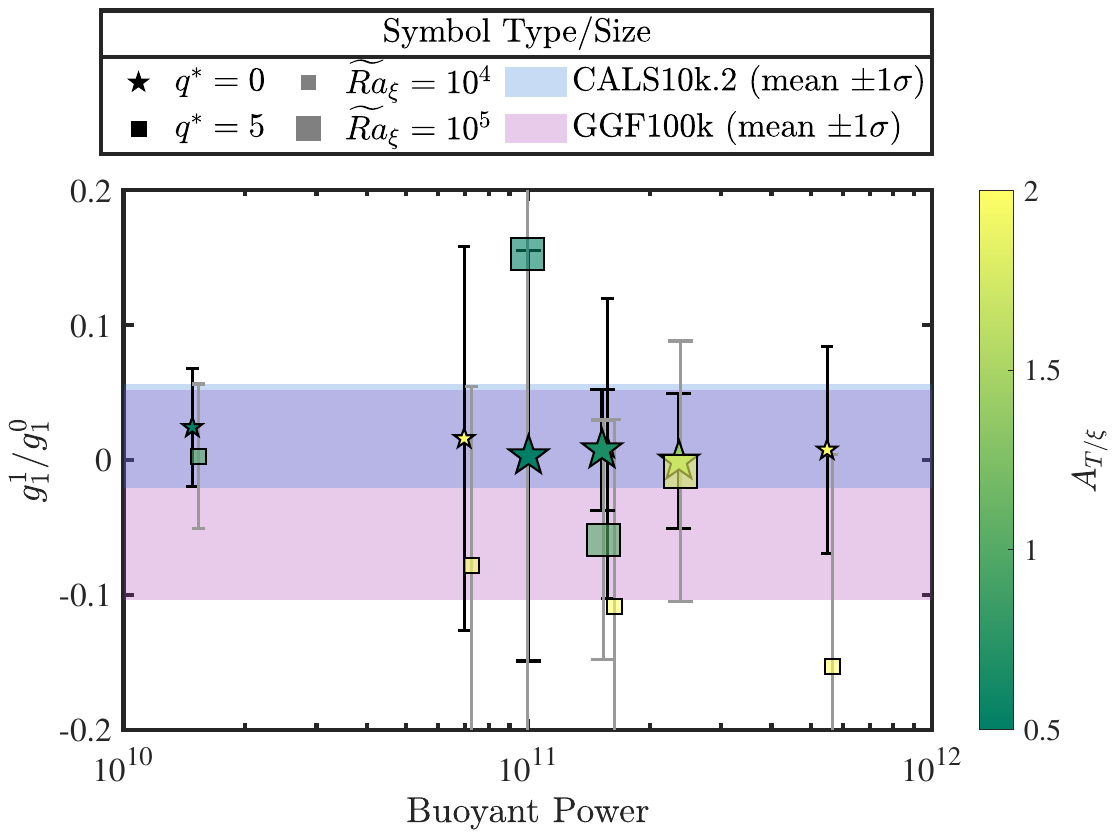}    
\put(-2,74){$(a)$}
\end{overpic}
\\
\hspace{21mm}
\begin{overpic}[width=0.86\linewidth,trim={0cm 6cm 0cm 11.2cm},clip]{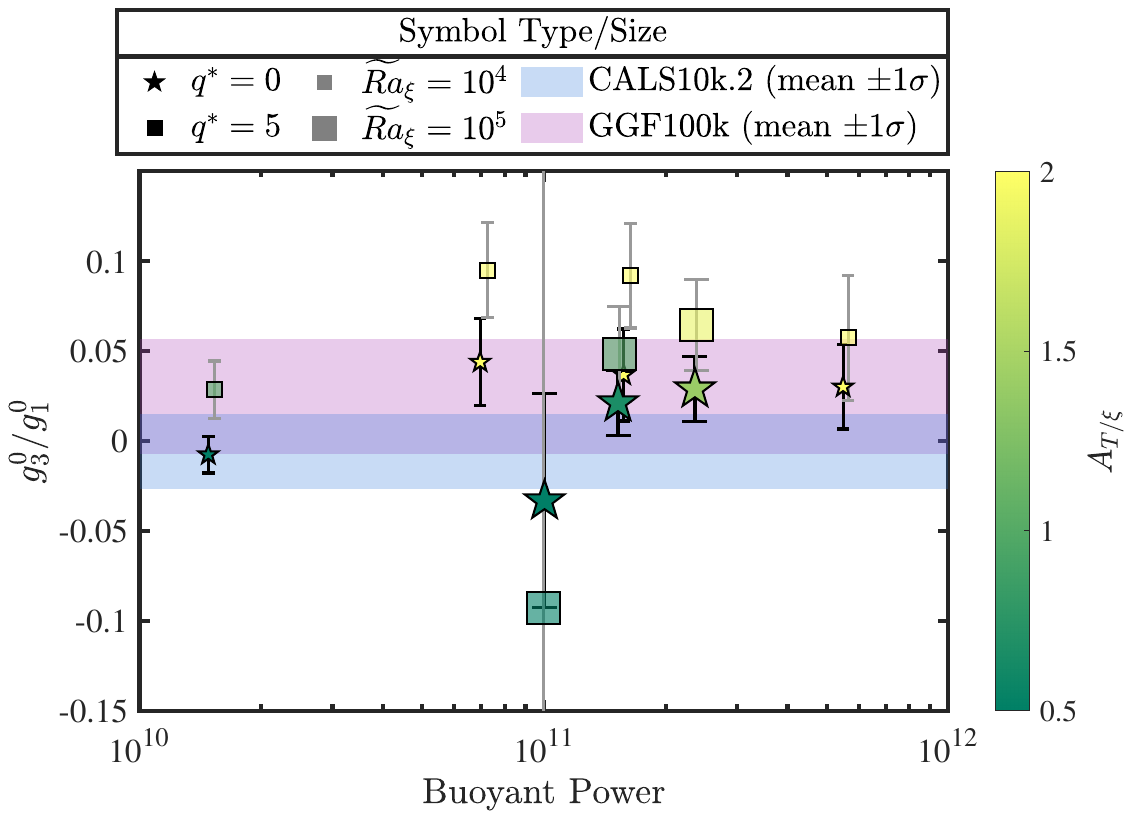}
\put(2,60){$(b)$}
\end{overpic}
\caption{Normalized Gauss coefficients (a) $\GC{1}{1}$ and (b) $\GC{3}{0}$ for all simulations compared with field models. Symbol shapes and sizes have the same meaning as in Figure \ref{fig:f_dip}. The symbol position represents the mean while the errorbars indicate one standard deviation in the time series of the Gauss coefficient ratios. Black (grey) errorbars are used for homogeneous (heterogeneous) cases. The observational ranges are indicated by horizontal color bands.}\label{fig:gcratio}
\end{figure}

Long-term observational models (e.g., GGF100k, CALS10k.2) are constrained for low spatial resolutions; we therefore compare the simulations and field models in terms of the low-degree ($\ell \leq 4$) time-averaged Gauss coefficients (normalised by $\gc{1}{0}$). The observational models are consistent with a zero mean for the non-dipole coefficients considered, although GGF100k displays more variability (Figure \ref{fig:gcratio}). Heterogeneous simulations tend to have non‑zero mean values, whereas the values for homogeneous cases approach zero (e.g., Figure \ref{fig:gcratio}a). That is, CMB heterogeneity tends to induce higher dipole tilt than the homogeneous cases, although a zero‑tilt scenario always lies within one standard deviation of the simulation average. 
The axial octupole component ($\GC{3}{0}$) has a larger magnitude in heterogeneous simulations than its homogeneous counterparts. Simulations with higher chemical buoyancy forcing (i.e., lower $\FACT$) tend to produce smaller values of $\GC{3}{0}$ than simulations with comparable buoyant power (Figure~\ref{fig:gcratio}b), and thus are more likely to overlap with the observational ranges.

\begin{figure}[h!]
\centering
\begin{overpic}[width=0.354\linewidth,trim={2cm 6cm 4cm 8cm},clip]{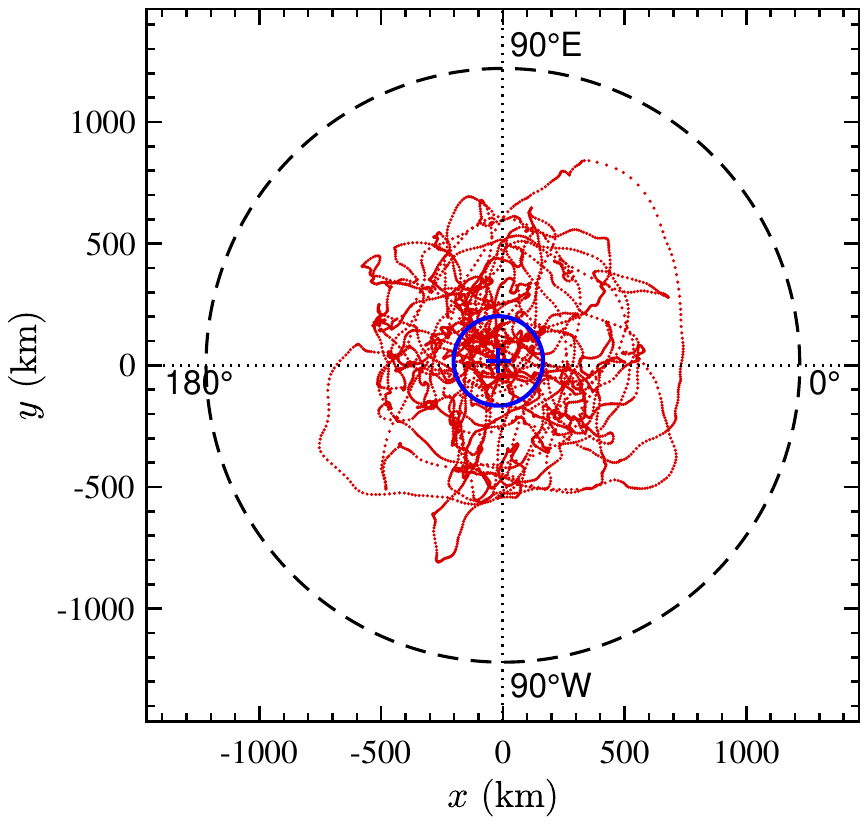}
\put(-2,95){$(a)$}
\end{overpic}
\begin{overpic}[width=0.318\linewidth,trim={3.5cm 6cm 4cm 8cm},clip]{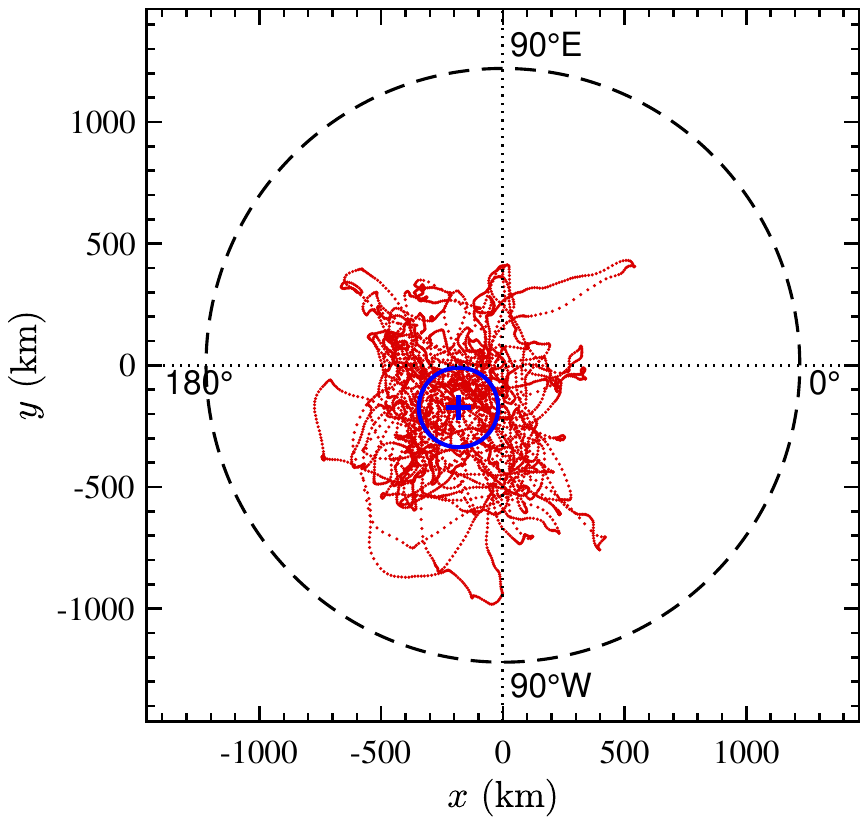}
\put(-2,95){$(b)$}
\end{overpic}
\begin{overpic}[width=0.318\linewidth,trim={3.5cm 6cm 4cm 8cm},clip]{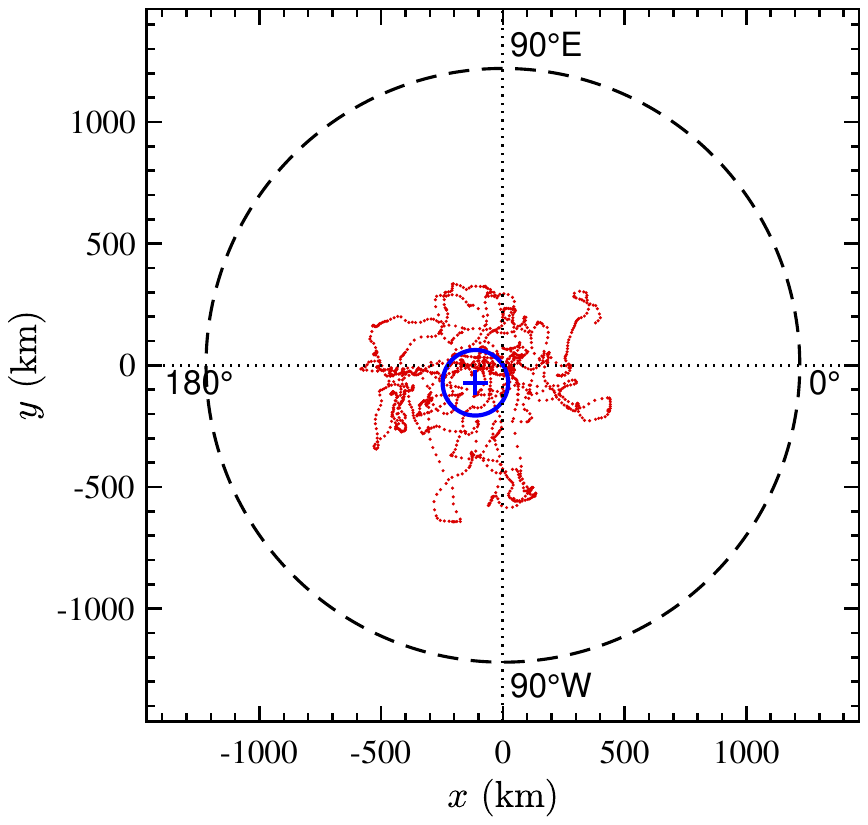}
\put(-2,95){$(c)$}
\end{overpic}
\caption{Timeseries of the radial position of the dipole at the equatorial plane for the three example cases in Figure~\ref{fig:TAF_map}, (i.e. $\tRaT=550$ with (a) $\qstar=0$, $\tRaC=10^{4}$ (b) $\qstar=5$, $\tRaC=10^{4}$ and (c) $\qstar=5$, $\tRaC=10^{5}$), calculated following \citet{james_1967}. The blue plus sign marks the time-averaged location, with the surrounding circle indicating one standard deviation. The black dashed circle indicate the inner core radius.}\label{fig:ecc_dipole}
\end{figure}

Given the tendency of heterogeneous simulations to induce dipole tilt, we also examine whether the dipole remains concentric with the rotation axis. The dipole eccentricity ($r_{dip}$) is generally higher for the heterogeneous dynamos than in their homogeneous counterpart (e.g., Figure~\ref{fig:ecc_dipole}) and tends to be offset towards the Pacific hemisphere. However, the mean eccentricity of the dipole across all simulations remains within the observational range, defined by the standard deviation of $r_{dip}$ in the field models (Figure \ref{fig:eccentricity}), and no systematic relationship is found between the dipole eccentricity and the thermal‑to‑chemical buoyancy ratio.

\begin{figure}[h!]
\centering
\begin{overpic}[width=0.47\linewidth,trim={0cm 0cm 0cm 15cm},clip]{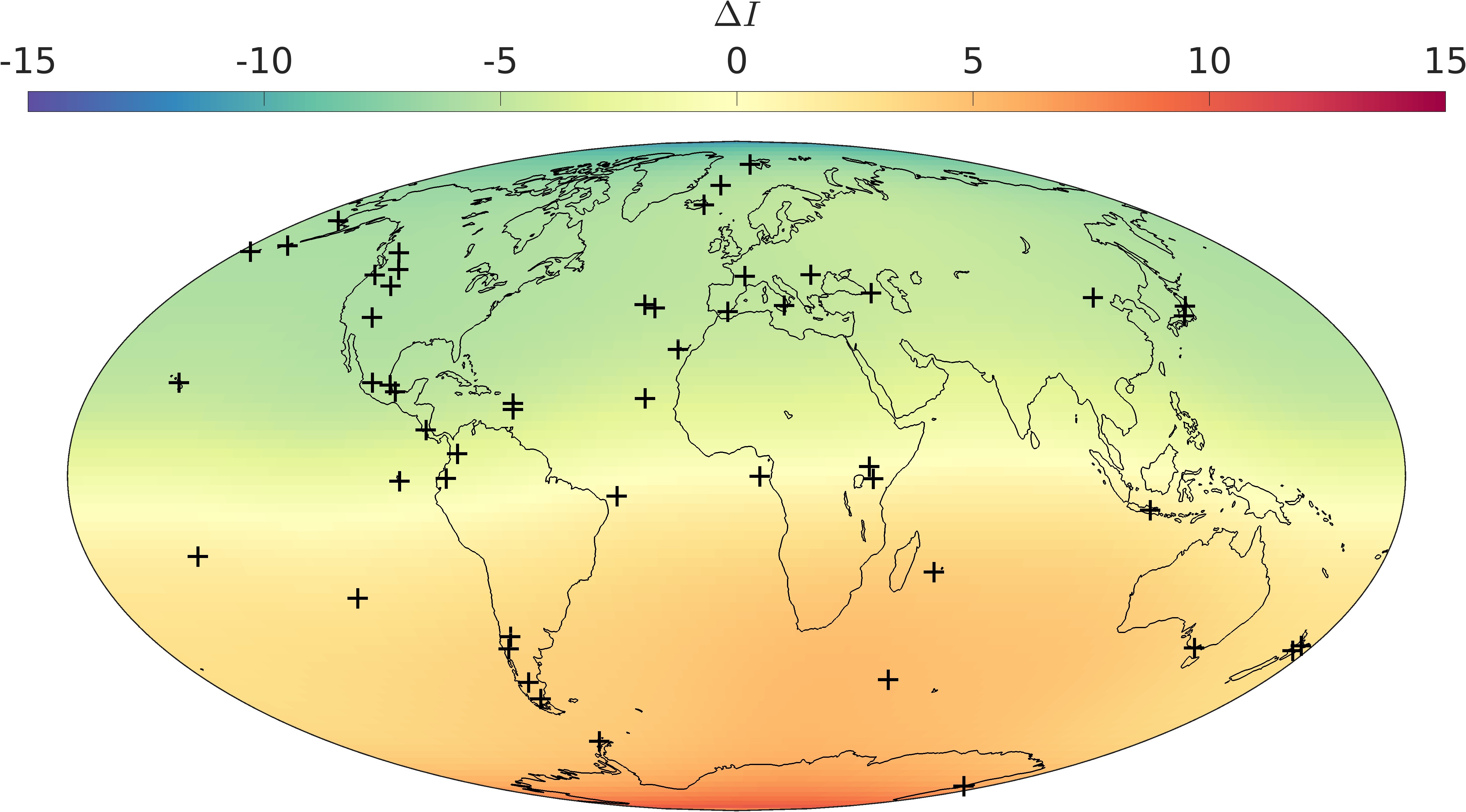}
\put(-2,55){$(a)\, \qstar=0,\ \tRaC=10^{4}$}
\end{overpic}
\begin{overpic}[width=0.47\linewidth,trim={0cm 0cm 0cm 0cm},clip]{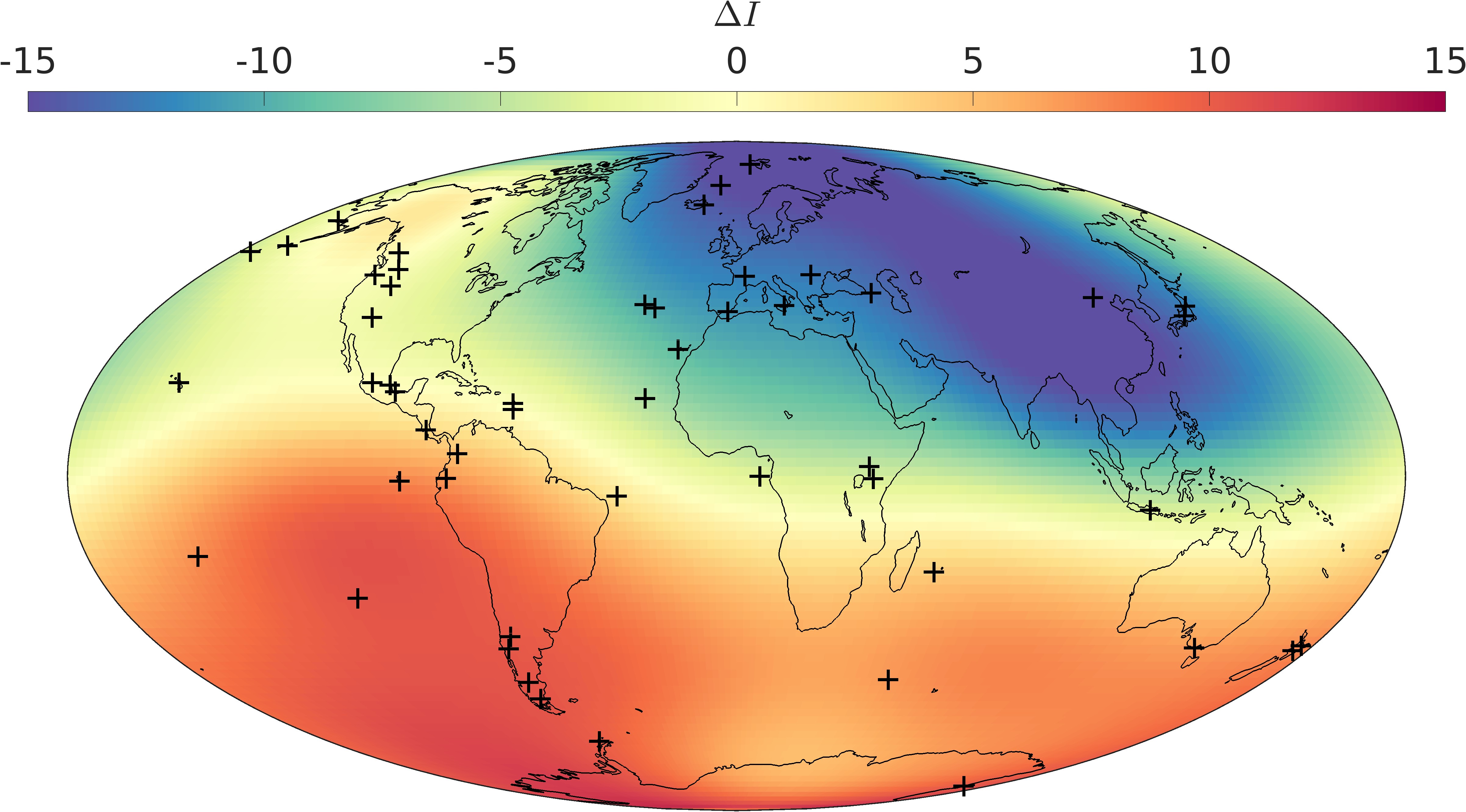}
\put(-2,55){$(b)\, \qstar=5,\ \tRaC=10^{4}$}
\end{overpic}
\vspace{8mm}

\begin{overpic}[width=0.47\linewidth,trim={0cm 0cm 0cm 15cm},clip]{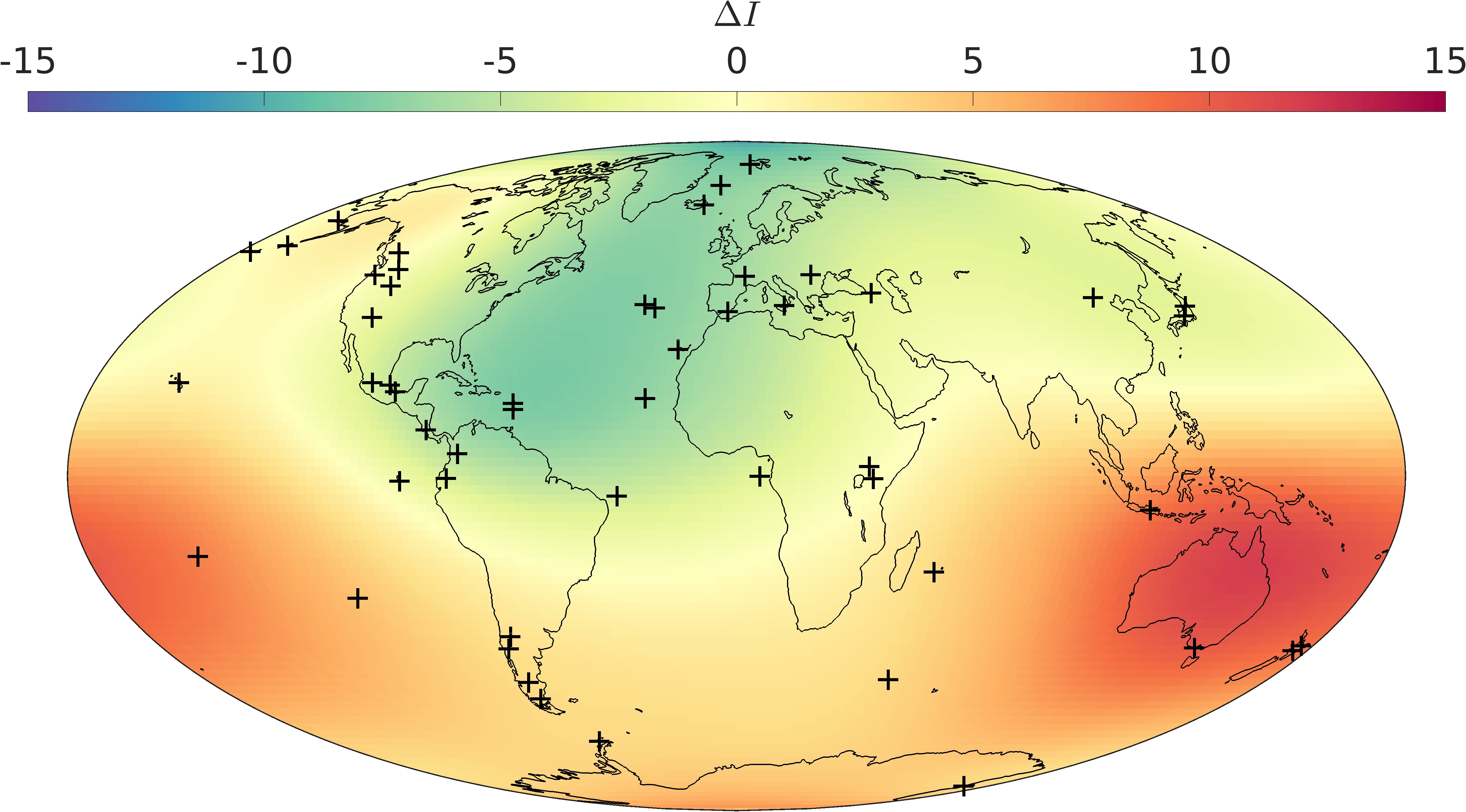}
\put(-2,50){$(c)\, \qstar=5,\ \tRaC=10^{5}$}
\end{overpic}
\begin{overpic}[width=0.47\linewidth,trim={0cm 0cm 0cm 15cm},clip]{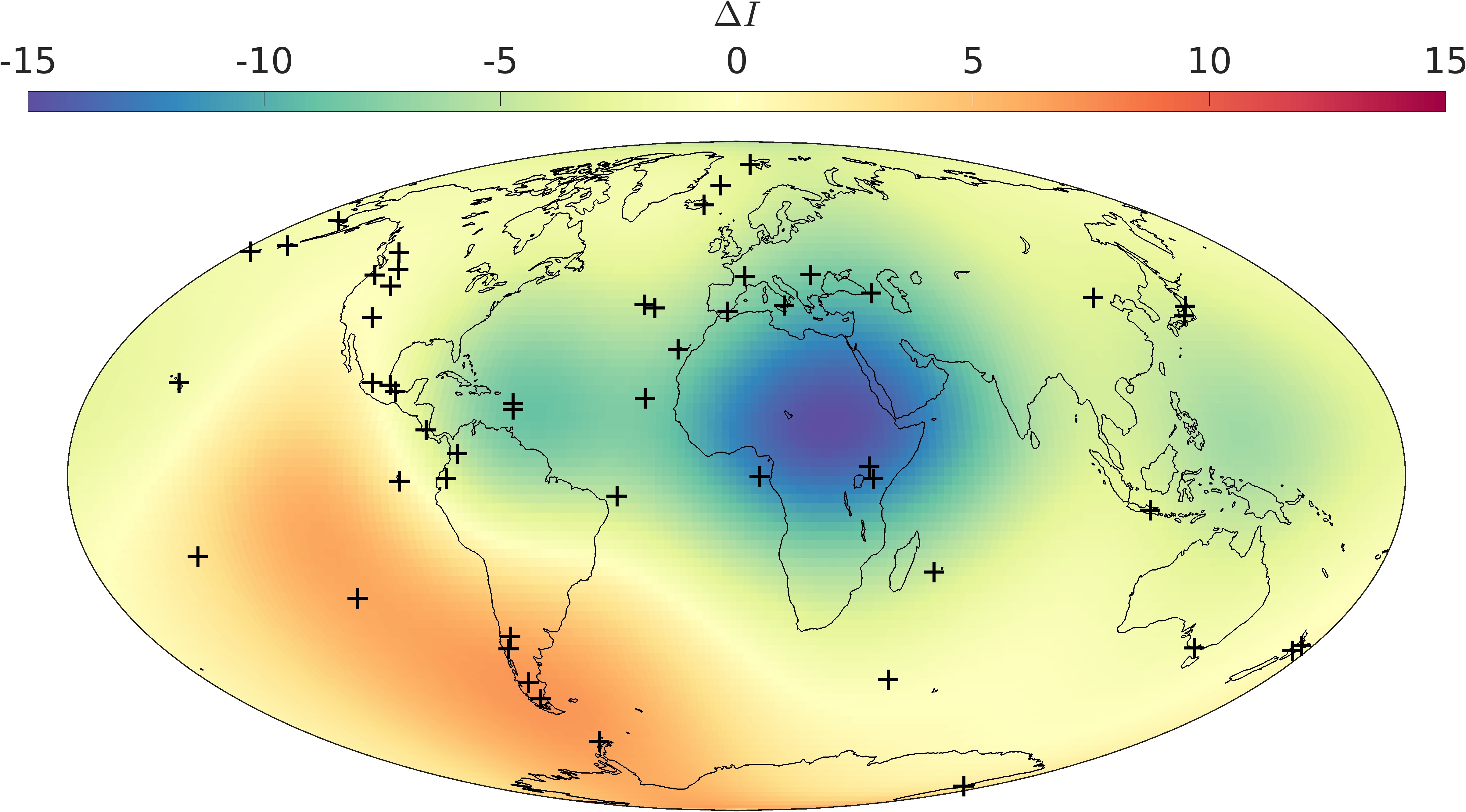}
\put(-2,50){$(d)$ GGF100k}
\end{overpic}
\caption{Time-averaged inclination anomaly ($\IncAnom$) maps at Earth's surface with site locations from the PSV10 database \citep{sprain_2019} marked by crosses. The time-averaged field has been truncated at $\truncation=10$.}\label{fig:inc_anom_map}
\end{figure}

\begin{figure}[h!]
\centering
\begin{overpic}[width=0.6\linewidth,trim={3cm 8cm 3cm 10cm},clip]{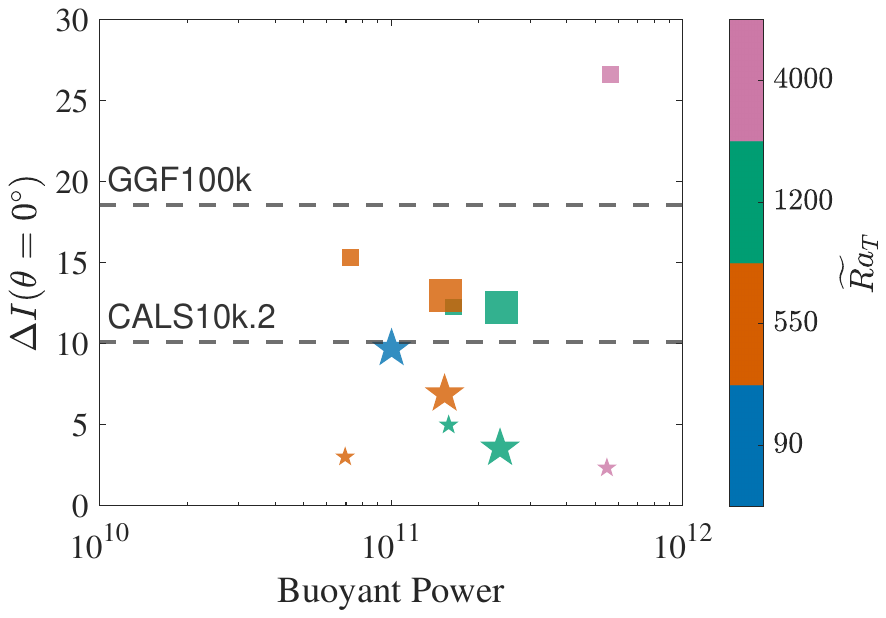}
\end{overpic}
\caption{Variation of time-averaged inclination anomaly range at the equator for the simulations in the range $0.35\leq\fdip\leq0.75$. The meaning of the symbols remains the same as Figure \ref{fig:f_dip}. The field at Earth's surface has been truncated at spherical harmonic degree $\truncation = 5$.}\label{fig:inc_anom_range}
\end{figure}

Among the quantities we examined for potential observable signatures of longitudinal field variation, the inclination anomaly ($\IncAnom$) provides clear and systematic behaviour. The time-averaged inclination anomaly maps of our three example cases (Figure \ref{fig:TAF_map}) are compared with the corresponding map from the GGF100k field model (Figure \ref{fig:inc_anom_map}). Uncertainties in field models arising from heterogeneous spatio-temporal sampling of the geomagnetic field together with limitations in the simulation's capacity to access the physical conditions of Earth's core mean that we do not expect simulated fields to reproduce the specific morphology of GGF100k. However, all homogeneous simulations (e.g., Figure~\ref{fig:inc_anom_map}a) fail to reproduce both the magnitude and the longitudinal variability evident in the observational model (Figure~\ref{fig:inc_anom_map}d). In contrast, heterogeneous simulations (e.g., \ref{fig:inc_anom_map}b,c) exhibit substantial longitudinal variability and a magnitude of $\IncAnom$ values that more closely resemble GGF100k. Quantifying the variability by the equatorial range of inclination anomaly (Figure~\ref{fig:inc_anom_range}), most homogeneous cases falls below the observational ranges, while heterogeneous cases readily reproduce anomaly ranges inferred from observations.  

\section{Discussion}\label{sec:discussion}
The most prominent structures induced by boundary heat-flux heterogeneity in our simulations appear to be the high-latitude flux patches (Figure~\ref{fig:TAF_map}b,c). The structures of the patches are broadly consistent with the simulation results reported by \citet{olson_2017}, although direct comparisons of the simulation outputs are difficult due to differences in setup. Nevertheless, we do not find a polar intensity maxima with a "lobe" structure of the radial field near the pole, as they reported for some cases (see Figure 7c in their paper). Rather, we find polar minima in the time averaged field (Figure~\ref{fig:TAF_map}) which deepen with increasing compositional forcing (Figure~\ref{fig:pol_min}) and boundary heterogeneity. We note that although our measure successfully reflects the TAF behaviour of the polar minima in our simulations, other measures \citep{cao_2018,lezin_2023} could be used to investigate the relative strength and persistence of the polar minima.

The non-dipolar fields (i.e., $\ell\ge2$) in our heterogeneous simulations tend to have significant non-zonal structure, with high-latitude patches of normal flux that persist at certain longitudes, similar to the findings of \citet[][]{biggin_2026}, although the details of the TAF morphology depends on the relative magnitude of buoyancy forcing (e.g., compare Figure~\ref{fig:br_ndip}b and c). 
Non-zonal structures at equatorial latitudes are also evident in our heterogeneous cases (Figure \ref{fig:Br_eq}), while such features are absent in the homogeneous cases, consistent with the findings of \citet{mound_2023}. In agreement with their simulations, our heterogeneous cases (e.g. Figure \ref{fig:Br_eq}b) exhibit a pair of reverse flux patches  straddling the equator, located roughly underneath South America. While this feature appears to be robust among our cases, an additional pair of similar patches, reported beneath the Indian Ocean, is only discernible in our thermally dominated cases. As with other TAF features, both the precise pattern and location of the equatorial patches depend on the balance of the buoyancy forcings.  The relative power in the zonal components in the non-dipolar field can be characterised by $\ZNDNDTAF$. However, this measure can exhibit slow convergence, depending on the simulation setup, and thus requires very long averaging time; therefore, it may be useful to look for measures that are easier to extract from the simulations. Additionally, a global field model is required to evaluate $\ZNDNDTAF$, and therefore it cannot be tested from local field measurements.


Paleo-secular variation at low latitude can be characterised by the angular dispersion of virtual geomagnetic poles ($S$), which can be parameterised using its median ($\Smed$) and interquartile range ($\Siqr$) \citep{biggin_2026}. Our heterogeneous setup closely resembles the {\em TCHET1} group of simulations of \citet{biggin_2026}, although we use a lower Ekman number. Our $\Siqr$ values display a broadly similar trend to their results (Figure~\ref{fig:s30}); however, over the parameter regime that we have explored, the contrast between homogeneous and heterogeneous simulations is less pronounced than reported by \citet[][]{biggin_2026}. We find that our homogeneous simulations yield a higher relative dipole strength than our heterogeneous cases, as quantified by the ratio $\ADNADTAF$, similar to their results.

Within our simulations, the influence of the thermal boundary condition is reflected in the higher inclination anomaly values observed in the heterogeneous cases compared to the homogeneous cases (Figure~\ref{fig:inc_anom_map}). The observational model GGF100k shows anomalies in the range ($-15^\circ\lesssim\IncAnom\lesssim 15^\circ$), with pronounced longitudinal variations. In contrast, the homogeneous simulations produce much weaker anomalies ($|\IncAnom|_{max}\lesssim8^\circ$) and lack longitudinal dependence. Although the pattern of inclination anomaly varies among the heterogeneous simulations, they consistently favour anomaly magnitudes comparable to GGF100k, along with a strong dependence on longitude (see also Figure S11 in \citet{biggin_2026}). This provides a systematic and observationally testable distinction between our homogeneous and heterogeneous simulations. \citet{davies_2008} also found strong longitudinal variations in inclination anomaly; however, their thermal setup uses low Rayleigh numbers to obtain weakly time-dependent solutions, compared to our vigourously turbulent simulations driven by high thermochemical forcing.

Another novel finding is the tilted eccentric dipole field produced by our heterogeneous simulations, compared to the nearly axisymmetric field produced in the homogeneous simulations. \citet{olson_2012} attributed a persistent eccentricity to lopsided inner core growth, whereas our simulations indicate that CMB heat flux heterogeneity can also lead to an eccentric dipole. Our heterogeneous simulations also favour a small dipole tilt (i.e., non-zero $\GC{1}{1}$ and $\HC{1}{1}$), unlike the dynamo reported in \citet{olson_2002} where the non-axial coefficients are not significant. \citet{davies_2008} employed a tomographic heat flux pattern similar to the one used in this study and found that the non-axisymmetric components increase with increasing degree of heterogeneity. Dipole tilt induced by heterogeneous CMB heat flux has been also reported by \citep{olson2018true}. However, the axial octupole ($\GC{3}{0}$) is the only coefficient in our simulations that is non-zero at the $1\sigma$ significance level, which holds for both homogeneous and heterogeneous simulations with $0.35\leq\fdip\leq0.75$. Previous studies \citep{christensen_2003,olson_2017, landeau_2017} have checked axisymmetric Gauss coefficients and also found the octupole to be the only significant component. Our simulations indicate an increase in $\GC{3}{0}$ when a heterogeneous CMB heat flux is imposed; conversely, $\GC{3}{0}$ decreases with increasing chemical buoyancy forcing ($\tRaC$).

The global field models, CALS10k.2 and GGF100k, suggest only a small non-zero $\GC{1}{1}$ for Earth's field over their respective time periods (i.e., zero is within one standard deviation of the observations, Figure~\ref{fig:gcratio}). Similarly, they have small octupole components (i.e., $\GC{3}{0}$ within one standard deviation of zero). Furthermore, the field model CALS10k.2 indicates a dipole eccentricity in the range $20\text{---}200$ km, whereas a larger range ($0\text{---}300$ km) is supported by GGF100k (Figure~\ref{fig:eccentricity}). These observational models provide a constraint on the plausible strength and variability of non-GAD features in Earth's field, but are limited to the most recent past. The slow evolution of mantle convection implies that the non-GAD features induced by the present-day CMB heat flux heterogeneity might also be probed by paleomagnetic data spanning millions of years \citep[see, e.g.,][]{olson_2012, biggin_2026}.

The heterogeneous heat flux conditions in our simulations induce stably stratified regions beneath Africa and the Pacific. For $\tRaC = 10^{5}$, these thermal regional inversion lenses \citep{mound_2019} can coexist with chemically stably stratified regions near the poles. Comparing our geodynamo simulations with our earlier convection runs at the same parameters \citep{naskar_2026}, we find very similar strength, thickness, and morphology of these thermal and chemical RILs (see Figure~\ref{fig:str_thk_RIL}), indicating limited influence of the magnetic field on these structures in the investigated parameter regime. Therefore, it seems that a good estimate of the properties of the RILs can be obtained from convection simulations, without resorting to expensive dynamo runs. Overall, we find that regions of stratified fluid a few hundred kilometres thick at the top of the core are a common feature of simulations in the region of parameter space that we have explored. The preferential LEA in polar regions found within our thermochemical simulations might provide a more visible seismic target than the thermally dominated RILs at equatorial regions.

\section{Conclusions}\label{sec:conclusion}
We have analysed $14$ new thermochemical geodynamos, $7$ homogeneous and $7$ heterogeneous simulations, to look for signatures of CMB heat flux heterogeneity, as well as the effect of top-heavy two-component convection dynamics on the simulated magnetic fields. All simulations display a QG-MAC force balance, with varying relative contributions from thermal and chemical buoyancy. We tested the simulations against existing criteria, assessing the semblance of the simulations with the geomagnetic field over a range of timescales.
Strong CMB heat flux heterogeneity induces thermal RILs beneath the African and Pacific LLVPs and strong chemical driving results in LEA preferentially in polar regions.

The key signature of RILs in the simulations is the longitudinal structure they induce in the long-term morphology and variability of the magnetic field. For example, the inclination anomaly of our heterogeneous simulations can reproduce both the magnitudes of $\IncAnom$ and the substantial longitudinal variations seen in observational models, whereas our homogeneous simulations cannot. We note that  $\IncAnom$ can be directly measured from spot readings of the field \citep[see e.g.][]{laj_2011} without the need of a global field model, making our inference easily testable. Our heterogeneous simulations also produce a tilted eccentric dipole, in contrast to the nearly axisymmetric fields produced by the homogeneous simulations, consistent with previous studies \citep{olson_2012,olson2018true}.

Our simulations prefer a non-zero axial octupole $\GC{3}{0}$, with this component being larger for heterogeneous models than homogeneous models, but generally decreasing in magnitude when $\tRaC$ is increased. All of our simulations also have polar minima in $B_r$, which deepen as the relative importance of chemical driving is increased. Increased chemical driving also promotes LEA in polar regions. Assessing the polar minima, to constrain relative thermal to chemical driving and hence the likelihood of significant LEA in polar regions, would benefit from good data coverage at high latitudes.

We have compared the thermal and chemical fields of the present geodynamo simulations with those of our previous non-magnetic simulations \citep{naskar_2026} with otherwise identical parameters. The thermally stratified RILs in the equatorial regions beneath Africa and the Pacific are robust features across all heterogeneous simulations, with properties and morphologies similar to those of the non-magnetic simulations. The chemically stratified RILs at the polar regions, driven by LEA, are also present and exhibit characteristics similar to those in non-magnetic simulations. 

In summary, geomagnetic features in our simulations depend on both the imposed CMB heat flux pattern and the relative importance of thermal to chemical buoyancy. CMB heterogeneity tends to promote non-GAD features, and in particular, longitudinally varying magnetic features. Increasing the importance of chemical buoyancy can suppress non-zonal magnetic features in our simulations, and promote the deepening of polar minima in $B_r$. Future simulations exploring a wider range of parameters (e.g., $\qstar$ and $\FACT$) and more extreme conditions (e.g., lower $\Ek$, greater separation between $\PrT$ and $\PrC$) should shed more light on both the degree of mantle control on and the importance of two-component convection in the dynamics of the core.     

\section*{CRediT authorship contribution Statement}\label{sec:credit}
\textbf{Souvik Naskar:} Writing - original draft, review \& editing, Visualisation, Validation, Software, Methodology, Formal Analysis, Conceptualisation. \textbf{Jonathan E. Mound:} Writing - review \& editing, Supervision, Formal Analysis, Project administration, Funding acquisition, Conceptualisation. \textbf{Christopher J. Davies:} Writing - review \& editing, Supervision, Formal Analysis, Project administration, Funding acquisition, Conceptualisation. \textbf{Hannah F. Rogers} Formal analysis, Data curation, Methodology.  \textbf{Stephen J. Mason} Formal analysis, Data curation, Methodology. \textbf{Andrew T. Clarke:} Software, Methodology, Resources.

\section*{Declaration of Competing interest}\label{sec:COI}
The authors declare that they have no competing financial interests or personal relationships that could have influence the work reported in this paper.

\section*{Acknowledgements}\label{sec:acknowledgements}
We gratefully acknowledge support from Natural Environment Research Council grants NE/W005247/1 (supporting SN, CJD and JEM), NE/Y003500/1 (supporting CJD and SJM), NE/V010867/1 (supporting CJD, ATC and HFR). This work used the \href{http://www.archer2.ac.uk}{ARCHER2 UK National Supercomputing Service}, and \href{https://arcdocs.leeds.ac.uk/welcome.html} {Aire}, part of the High-Performance Computing facilities at the University of Leeds, UK.

\appendix

\section{Supplementary Information}\label{app:SI}
\setcounter{figure}{0}
\renewcommand{\thefigure}{A\arabic{figure}}

\setcounter{table}{0}
\renewcommand{\thetable}{A\arabic{table}}

\setcounter{equation}{0}
\renewcommand{\theequation}{A\arabic{equation}}
\input{supplementary_arXiv}

\section{Simulation Diagnostics}\label{app:diagnostics}
Additional data is available in the supplementary file \href{https://shorturl.at/EANnT}{\texttt{supplementary\_data.xlsx}} included with this submission.

\bibliographystyle{elsarticle-harv}
\bibliography{elsarticle}

\end{document}

%% file: supplementary_arXiv.tex
\subsection{ Detailed Methodology}\label{sec:methods_sup}
The governing equations for the conservation of mass, momentum, magnetic field, energy, and chemical composition can be written as follows:

\begin{equation}\label{eqn:continuity_d}
   \boldsymbol{\nabla\cdot u}=\boldsymbol{\nabla\cdot B}=0
\end{equation}

\begin{equation}\label{eqn:momentum_d}
\begin{split} 
\left(\frac{\partial \boldsymbol{u}}{\partial t}+    (\boldsymbol{u}\cdot\boldsymbol{\nabla})\boldsymbol{u}\right)+2(\boldsymbol{\Omega}\times\boldsymbol{u})  =-\boldsymbol{\nabla}P+\frac{\trho}{\rho_o}\boldsymbol{g}+\frac{1}{\rho_o\mu_o}\left(\boldsymbol{\nabla}\times\boldsymbol{B}\right)\times \boldsymbol{B}+
\nu{\nabla^{2}}\boldsymbol{u}
\end{split}
\end{equation}

\begin{equation}\label{eqn:maxwell_d}
\frac{\partial \boldsymbol{B}}{\partial t} =\boldsymbol{\nabla}\times(\boldsymbol{u}\times \boldsymbol{B})+
\eta{\nabla^{2}}\boldsymbol{B}
\end{equation}

\begin{equation}\label{eqn:energy_d}
\frac{\partial T}{\partial t}+    (\boldsymbol{u}\cdot\boldsymbol{\nabla})T =
\kappa_{T}{\nabla^{2}}T+S_T
\end{equation}

\begin{equation}\label{eqn:composition_d}
\frac{\partial \xi}{\partial t}+    (\boldsymbol{u}\cdot\boldsymbol{\nabla})\xi =
\kappa_{\xi}{\nabla^{2}}\xi+S_{\xi}
\end{equation}

The modified pressure, $P$, includes the centrifugal potential. Gravity varies linearly with the radius such that $\boldsymbol{g}= -(g_o/r_o)\boldsymbol{r}$, where $g_o$ is the reference gravitational acceleration at CMB ($r=r_o$). We use the Boussinesq approximation where the variations of the density $\trho$ due to the temperature $\tT$ and concentration of light elements $\tC$ are only taken in the buoyancy force. Following a linear equation of state, we get, 

\begin{equation}\label{eqn:bousensq}
    \frac{\trho}{\rho_o}=\frac{\rho-\rho_o}{\rho_o}=-\alpha_T(\tT-T_o)-\alpha_\xi(\tC-\xi_o),
\end{equation}
by assuming $|\rho-\rho_o|/\rho_o \ll 1$, where $T_o,\xi_o,\rho_o$ are the reference values at $r=r_o$. Because the scalar magnitudes have no dynamic significance (unlike the gradients of these scalar quantities), we set $(T_o, \xi_o) = (0, 0)$. In this equation of state \ref{eqn:bousensq}, $\tT$ is the departure from the destabilising background superadiabatic temperature profile ($T_c$, henceforth referred to as the conductive state), whereas $\tC$ is the departure from the compositional reference barodiffusive profile $\xi_c$ \citep{davies_2011}, as expressed below. 

\begin{equation}\label{eqn:tempdef}
    \begin{split}
    \tT=T-T_c \\
    \tC=\xi-\xi_c
    \end{split}
\end{equation}

Fixed-flux thermal boundary conditions are imposed at the inner and outer boundaries such that the total radial heat flow is equal through the inner and outer surfaces ($\boldsymbol{Q}_{T,i}=\boldsymbol{Q}_{T,o}$). The temperature gradient at the boundaries is expressed as $\boldsymbol{\nabla}T_{c}=-(\beta_{T}/r^{2})\boldsymbol{\hat{r}}$ and related to the heat flux at the inner core boundary (ICB) through Fourier Law as, $\boldsymbol{Q}_{T,i}=4\pi r_{i}^{2}\boldsymbol{q_{T}^{avg}}=4\pi r_{i}^{2}(-k_{T}\boldsymbol{\nabla}T_{c})=4\pi k_{T} \beta_T \boldsymbol{\hat{r}}$. The thermal convection is maintained entirely by this heat flow, and therefore, we set $S_T=0$ in equation  \ref{eqn:energy_d}. Without a heat source, the thermal conduction equation can be expressed as,

\begin{equation}\label{eqn:conduction}
\frac{\kappa_T}{r^2}\frac{d}{dr} \left( r^2\frac{dT_c}{dr} \right) = 0
\end{equation}
subjected to the boundary conditions,
\begin{equation}\label{eqn:conduction_bc}
\left( \frac{dT_c}{dr} \right)_{r=r_i} = -\frac{\beta_T}{r_i^{2}},\qquad
\left( \frac{dT_c}{dr} \right)_{r=r_o} = -\frac{\beta_T}{r_o^{2}}
\end{equation}

The solution is,
\begin{equation}\label{eqn:conduction_sol}
\frac{dT_c}{dr} = -\frac{\beta_T}{r^{2}},\qquad
T_c(r) = \frac{\beta_T}{r}+c_1
\end{equation}

Here, the constant of integration $c_1$ is not constrained by the boundary conditions. It can be set to zero without any loss of generality, as only the gradient has dynamic significance. The temperature drop across the shell for purely thermal conduction will be,

\begin{equation}\label{eqn:tempdrop}
\Delta T_c= \beta_T\left(\frac{1}{r_i}-\frac{1}{r_o}\right)=\frac{\beta_Th}{r_ir_o}
\end{equation}
where $h=r_o-r_i$ is the shell gap.  

We assume zero compositional flux at the CMB (i.e. $\boldsymbol{{Q}}_{\xi,o}=\boldsymbol{0}$), while imposing fixed flux conditions at the inner boundary. To ensure stationary solutions, we assume that the flux from the inner core is balanced by a spatially homogeneous sink ( $S_\xi$ ) that maintains the global balance of lighter elements. With a sink term, the light element diffusion equation can be expressed as,

\begin{equation}\label{eqn:massdiffusion}
\frac{\kappa_\xi}{r^2}\frac{d}{dr} \left( r^2\frac{d\xi_c}{dr} \right) +S_\xi=0
\end{equation}

subjected to the boundary conditions,

\begin{equation}\label{eqn:massdiff_bc}
\left( \frac{d\xi_c}{dr} \right)_{r=r_i} = -\frac{\beta_\xi}{r_i^{2}},\qquad
\left( \frac{d\xi_c}{dr} \right)_{r=r_o} = 0
\end{equation}

The solution is,

\begin{equation}\label{eqn:massdiff_sol}
\frac{d\xi_c}{dr} = \frac{\beta_\xi}{r_o^3-r_i^3}(r-r_o^3/r^2),\qquad
\xi_c(r) = \frac{\beta_\xi}{2(r_o^3-r_i^3)}(r^2+2r_o^{3}/r)+c_2
\end{equation}

where, $S_\xi=-3\beta_\xi\kappa_\xi/(r_o^{3}-r_i^3)$, and the constant of integration $c_2$ is not constrained by the boundary conditions. The boundaries are electrically insulated.

\subsubsection{Non-dimensional governing equations}\label{sec:gov_eq_nd}

We proceed by non-dimensionalising equations \ref{eqn:continuity_d}-\ref{eqn:composition_d}  using the shell gap $h$ as length scale, the magnetic diffusion time, $h^2/\eta$ , as time-scale, $\beta_T/h$ as temperature scale, $\beta_\xi/h$ as the compositional field scale and $(2\Omega\rho_o\mu_o\eta)^{1/2}$ as the scale for the magnetic field. For these choices, the non-dimensional equations become,
\begin{equation}\label{eqn:continuity_nd}
   \boldsymbol{\nabla^{*}\cdot u^{*}}=\boldsymbol{\nabla^{*}\cdot B^{*}}=0
\end{equation}

\begin{equation}\label{eqn:momentum_nd}
\begin{split} 
\frac{E}{Pm}\left(\frac{\partial \boldsymbol{u^{*}}}{\partial t^{*}}+    (\boldsymbol{u^{*}}\cdot\boldsymbol{\nabla^{*}})\boldsymbol{u^{*}}\right)+\boldsymbol{\hat{z}}\times\boldsymbol{u^{*}}  =-\boldsymbol{\nabla^{*}}P^{*}+\\
Pm\left(\frac{\widetilde{Ra}_{T}}{Pr_T}\tT^{*}+\frac{\widetilde{Ra}_{\xi}}{Pr_\xi}\tC^{*}\right)\boldsymbol{r}^{*}+(\boldsymbol{\nabla}\times\boldsymbol{B^{*}})\times \boldsymbol{B^{*}}+
E{\nabla^{*2}}\boldsymbol{u^{*}}
\end{split}
\end{equation}

\begin{equation}\label{eqn:maxwell_nd}
\frac{\partial \boldsymbol{B^{*}}}{\partial t^{*}} = \boldsymbol{\nabla}\times(\boldsymbol{u^{*}}\times \boldsymbol{B}^{*})+{\nabla^{*2}}\boldsymbol{B^{*}}
\end{equation}
 
\begin{equation}\label{eqn:energy_nd}
\frac{Pr_{T}}{Pm}\left(\frac{\partial T^{*}}{\partial t^{*}}+    (\boldsymbol{u}^{*}\cdot\boldsymbol{\nabla}^{*})T^{*}\right) = {\nabla^{*2}}T^{*}
\end{equation}
 
\begin{equation}\label{eqn:composition_nd}
\frac{Pr_{\xi}}{Pm}\left(\frac{\partial \xi^{*}}{\partial t^{*}}+    (\boldsymbol{u}^{*}\cdot\boldsymbol{\nabla}^{*})\xi^{*}\right) =
{\nabla^{*2}}\xi^{*}-\frac{3}{r^{*3}_o-r^{*3}_i}
\end{equation}

These equations are subjected to no-slip boundary conditions for the velocity at both boundaries. Also, the modified flux Rayleigh number used in this study relates to the flux Rayleigh number as $\tRaT=Ra_TE$ (and similarly $\tRaC=Ra_\xi E$). In what follows, and throughout the main text, the superscript asterisk is omitted, and all variables are understood to be non‑dimensional unless stated otherwise.

\subsubsection{Non-dimensional background states}\label{sec:background_nd}

Non-dimensionalising the background profiles in equations \ref{eqn:conduction_sol} and \ref{eqn:massdiff_sol}, we get,

\begin{equation}\label{eqn:conduction_sol_nd}
\frac{dT_c}{dr} = -\frac{1}{r^{2}},\qquad
T_c(r) = \frac{1}{r}+c_1,\qquad
\end{equation}

\begin{equation}\label{eqn:conduction_temp_diff}
\Delta T_c = T_c(r_i)-T_c(r_o)=\frac{1}{r_i}-\frac{1}{r_o}=\frac{(1-\chi)^{2}}{\chi}
\end{equation}
where $\chi=r_i/r_o$ is the radius ratio fixed at $\chi=0.35$ in our study.
\begin{equation}\label{eqn:diff_profile_nd}
\begin{split}
\frac{d\xi_c}{dr} = \frac{1}{r_o^{3}-r_i^{3}}(r-r_o^{3}/r^{2}),\\
\xi_c(r) = \frac{1}{2(r_o^{3}-r_i^{3})}(r^{2}+2r_o^{3}/r)+c_2    
\end{split}
\end{equation}

\begin{equation}\label{eqn:diffusion_comp_diff}
\begin{split}
\Delta \xi_c= \xi_c(r_i)-\xi_c(r_o)=\frac{(1-\chi)^{2}}{\chi^2}\frac{\chi(\chi+2)}{2(1+\chi+\chi^{2})}    
\end{split}
\end{equation}


\subsection{Force balance}
\begin{figure}[h!]
\centering
\begin{overpic}[width=0.335\linewidth,trim={0cm 0cm 0cm 0cm},clip]{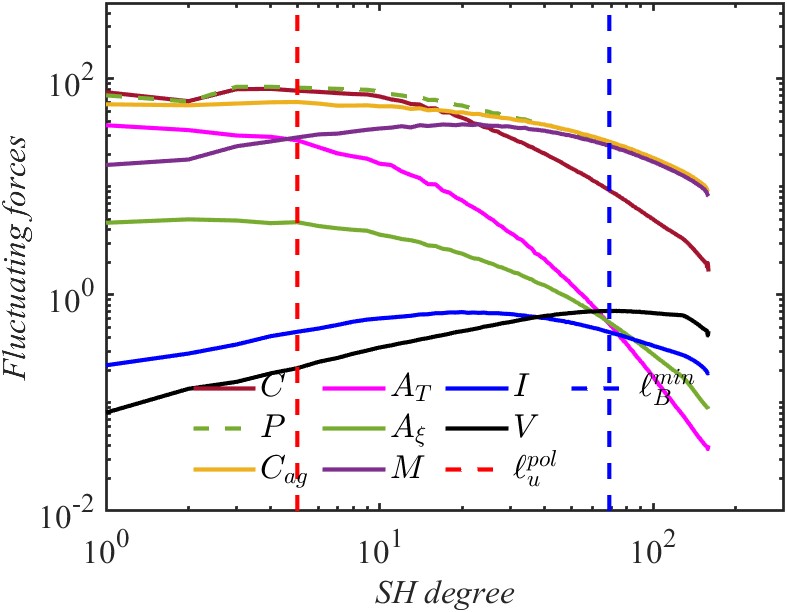}
\put(-2,80){$(a)$}
\end{overpic}
\begin{overpic}[width=0.32\linewidth,trim={1.2cm 0cm 0cm 0cm},clip]{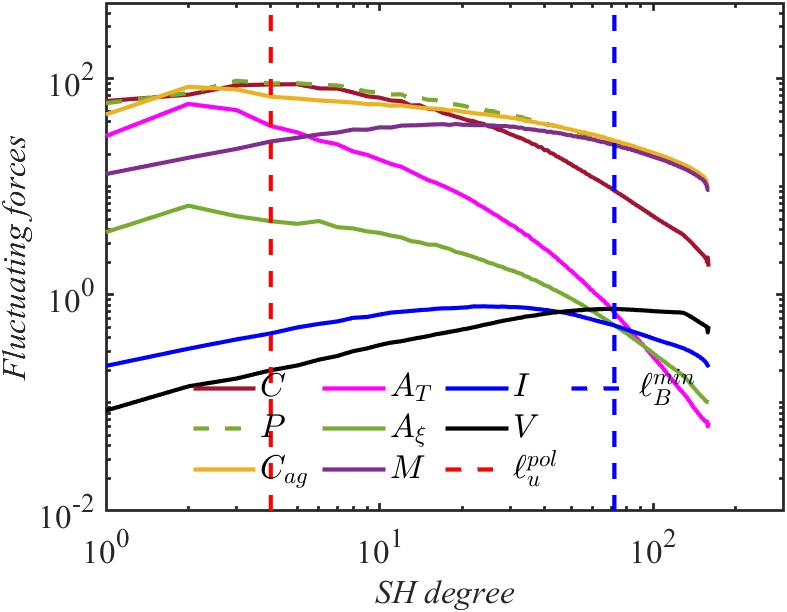}
\put(-2,80){$(b)$}
\end{overpic}
\begin{overpic}[width=0.32\linewidth,trim={1.2cm 0cm 0cm 0cm},clip]{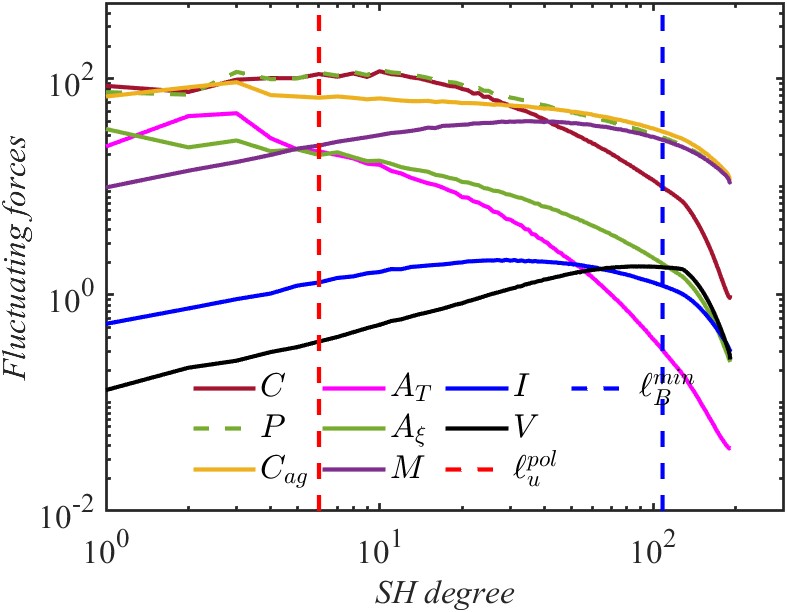}
\put(-2,80){$(c)$}
\end{overpic}
\caption{Force spectra for (a) homogeneous, $\tRaC = 10^{4}$ (b) heterogeneous, $\tRaC = 10^{4}$ and (c) heterogeneous, $\tRaC = 10^{5}$ at $\tRaT = 550$. The fluctuating forces defined in equation \ref{eqn:momentum_f} are averaged over the volume, excluding $10$ times the Ekman-layer depth adjacent to the upper and lower boundaries.}\label{fig:fspec}
\end{figure}

We refer to the terms from left to right in equation \ref{eqn:momentum_nd} as time-derivative ($\boldsymbol{TD}$), inertia ($\boldsymbol{I}$), Coriolis ($\boldsymbol{C}$), pressure gradient ($\boldsymbol{P}$), thermal buoyancy (or thermal Archemedian, $\boldsymbol{A_T}$), chemical buoyancy (or chemical Archemedian, $\boldsymbol{A_\xi}$), Lorentz (or Magnetic, $\boldsymbol{M}$) and the viscous ($\boldsymbol{V}$) forces. Additionally, the ageostrophic Coriolis force $\left(\boldsymbol{\hat{z}}\times\boldsymbol{u}\right)+\boldsymbol{\nabla}\widetilde{P}$ is denoted as $\boldsymbol{C_{ag}}$. We partition dependent variables into their azimuthally averaged mean and corresponding fluctuating parts as,
\begin{equation}\label{eqn:def_average}
\begin{split}
    f(r,\theta,\phi,t)=\overline{f}(r,\theta,t)+f '(r,\theta,\phi,t),\\
    \overline{f}(r,\theta,t)=\frac{1}{2\pi}\int_{0}^{2\pi}f(r,\theta,\phi,t)d\phi.
\end{split}    
\end{equation}
Azimuthally averaging equation \ref{eqn:momentum_nd}, and then substracting it from equation \ref{eqn:momentum_nd}, leads to the fluctuating momentum equation

\begin{align}\label{eqn:momentum_f}
&\underbrace{\frac{E}{Pm}\frac{\partial \boldsymbol{u'}}{\partial t}}_{TD'}+
\underbrace{\frac{E}{Pm}\{(\boldsymbol{\overline{u}}\cdot\boldsymbol{\nabla})\boldsymbol{u'}+    (\boldsymbol{u'}\cdot\boldsymbol{\nabla})\boldsymbol{\overline{u}}+    (\boldsymbol{u'}\cdot\boldsymbol{\nabla})\boldsymbol{u'}-(\overline{\boldsymbol{u'}\cdot\boldsymbol{\nabla})\boldsymbol{u'}}\}}_{I'}
+\underbrace{\left(\boldsymbol{\hat{z}}\times\boldsymbol{u'}\right)}_{C'}=
\underbrace{-\boldsymbol{\nabla}P'}_{P'}\notag \\
&+\underbrace{\Pm\frac{\tRaT}{\PrT}\tT'}_{A_T'}\boldsymbol{r}
+\underbrace{\Pm\frac{\tRaT}{\PrC}\tC'}_{A_\xi'}\boldsymbol{r}
+\underbrace{(\boldsymbol{\overline{B}}\cdot\boldsymbol{\nabla})\boldsymbol{B'}+    (\boldsymbol{B'}\cdot\boldsymbol{\nabla})\boldsymbol{\overline{B}}+    (\boldsymbol{B'}\cdot\boldsymbol{\nabla})\boldsymbol{B'}-(\overline{\boldsymbol{B'}\cdot\boldsymbol{\nabla})\boldsymbol{B'}}}_{M'}\notag \\
&+\underbrace{{E\nabla^{2}}\boldsymbol{u'}}_{V'},
\end{align}

where prime ($\prime$) denotes a fluctuating component. A detailed analysis of different methods for analysing dynamical balances in rotating spherical shell convection is given in \citet{naskar_2025b}. The scale-dependent fluctuating forces in our three example cases are presented in figure \ref{fig:fspec}.

\subsection{Global measures of field morphology and variability}\label{sec:compliance_SI}
\begin{figure}[h!]
\centering
\begin{overpic}[width=0.8\linewidth,trim={0cm 0cm 0cm 0cm},clip]{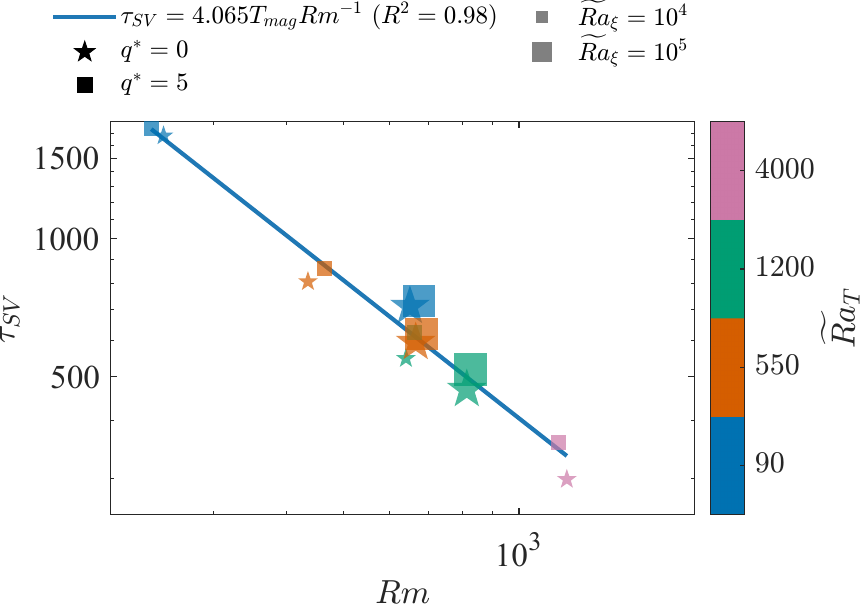}
\end{overpic}
\caption{The variation of the master time scale of secular variation with the magnetic Reynolds number. In the best-fit equation, $T_{mag}$ is the time-scale of magnetic diffusion.}\label{fig:Rm_vs_tau_SV}
\end{figure}

We have used four measures of compliance with the geomagnetic field as listed below.
\begin{enumerate}
    
    \item The temporal variability of the field is represented by the master secular variation timescale $\tausv$ \citep{lhuillier_2011,holme_2011}. Recent satellite models exhibit a value in the range $\tausv\simeq370-470\ \mathrm{yr}$ \citep{lhuillier_2011}. The variation of the master time scale of secular variation with the magnetic Reynolds number is presented in Figure \ref{fig:Rm_vs_tau_SV}.

    \item The semblance with the modern field of the past few centuries are accessed through the four compliance criteria proposed by \citet{christensen_2010}.
    \begin{enumerate}
        \item Relative axial dipole power is measured as the ratio of the axial dipole power to the non-axial dipole field, $\adnad$.
        \item Relative equatorial symmetry of the field measured by the ratio of spherical harmonic components with $(l+m)$ even (symmetric) to odd (antisymmetric), $\odev$.
        \item Relative zonality of the field measured by the ratio of zonal ($m=0$) to non-zonal ($m\geq1$) components of the non-dipolar field ($l\geq2$), $\znz$.
        \item Flux concentration factor measuring the relative variance in the squared radial field, $\fcf$.
    \end{enumerate}

  Each of these four measures ($C_i$) is evaluated as a degree of non-compliance of a simulation with a modern field model $gufm1$, and an estimate of variability ($\sigma_{i}$), such that $\chi^2_{i}= [\{ln(C^{sim}_{i})-ln(C^{gufm1}_{i})\}/ln(\sigma_{i})]^{2}$, and the total non-compliance ($\chisqtot=\sum\chi_{i}^2$). All the measures are summarized in \ref{app:diagnostics}, while the minimum value in the time series of total compliance ($\chisqtot$) is used in Figure \ref{fig:compliance_vs_fdip}(b).

  \item Another measure of variability is the paleo-secular variation index \citep{panovska_2017}, $P_i$, representing the departure of virtual geomagnetic poles from a geocentric axial dipole field, combined with the deviation of virtual dipole moment from the present day value. In figure \ref{fig:compliance_vs_fdip}(b), the time median of the global mean $P_i$ has been used where the values are downsampled at site locations of the paleomagnetic dataset PSV10 \citep{cromwell_2018} following  \citet{sprain_2019}. For reference, the index varies in the range $P_i\sim0.08-0.2$ in the field model GGF100k \citep{mason_2024}. 
  
  
  \item Compliance with paleomagnetic field for the last 10Ma is assessed using the diagnostics proposed by \citet{sprain_2019}.  These include the following.
  \begin{enumerate}
      \item The "Model G" coefficients $a$ and $b$ representing the scatter of virtual geomagnetic poles as a measure of paleosecular variation. 
      \item Inclination anomaly $\Delta I$ as a measure of time-averaged field behaviour.
      \item The ratio of the interqurtile range to the median of the virtual dipole moment, $V\%$, as a measure of temporal variation in magnetic intensity.
      \item A measure of transition time $\tau_t$ to represent the characteristic of the simulated field against geomagnetic field revarsals.
  \end{enumerate}
  The five criteria described above are estimated as a misfit ($\DelQPM^{i}$) from their corresponding Earth-like values \citep{sprain_2019} where the total misfit is calculated as $\DelQPM=\sum\DelQPM^{i}$ and presented in Figure \ref{fig:compliance_vs_fdip}(a).

\end{enumerate}

\clearpage

\subsection{Quantification of regional magnetic signatures}\label{sec:regional}

\begin{figure}[h!]
\centering
\begin{overpic}[width=0.8\linewidth,trim={2cm 9cm 2cm 9cm},clip]{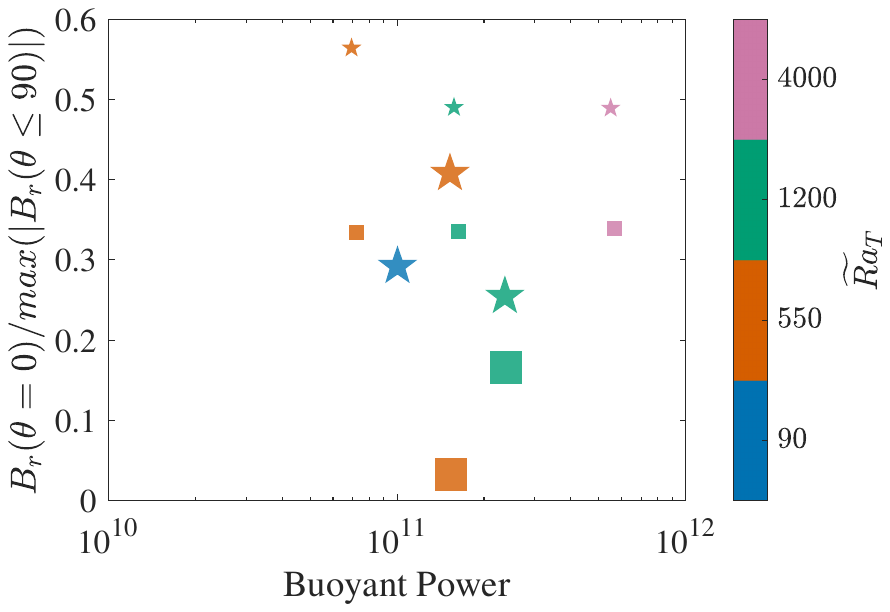}
\end{overpic}
\caption{The variation of the strength of the polar minima for the simulations with $0.35\leq \fdip \leq 0.75$. The time- and longitudinally averaged field has been used here. The symbols have the same meaning as in Figure \ref{fig:Rm_vs_tau_SV}.}\label{fig:pol_min}
\end{figure}

\begin{figure}[h!]
\centering
\begin{overpic}[width=0.8\linewidth,trim={0cm 6cm 1cm 7cm},clip]{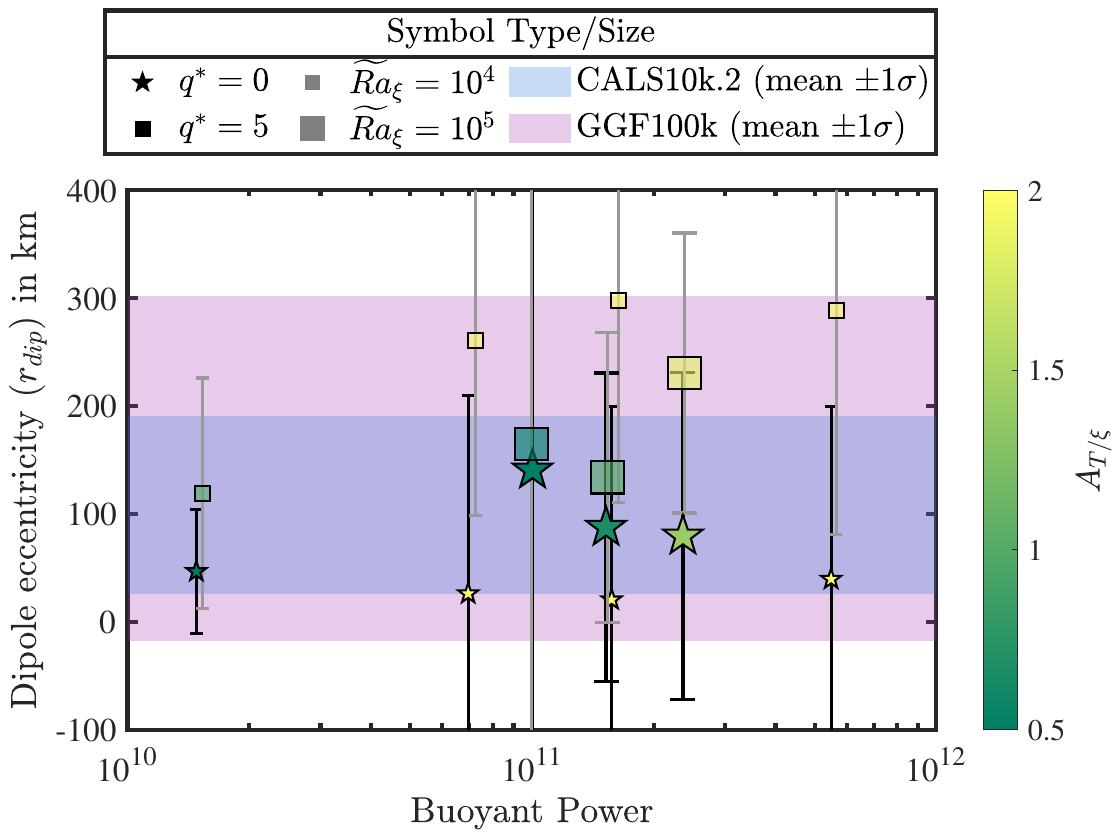} 
\end{overpic}
\caption{Distance of the centre of the dipole vector from Earth's centre in the equatorial plane. The symbols represent the mean values, while the errorbars represent the standard deviation.  Black (grey) errorbars are used for homogeneous (heterogeneous) cases. The observational ranges are indicated by horizontal color bands.}\label{fig:eccentricity}
\end{figure}


\begin{figure}[h!]
\centering
\begin{overpic}[width=0.7\linewidth,trim={1cm 9cm 2cm 7cm},clip]{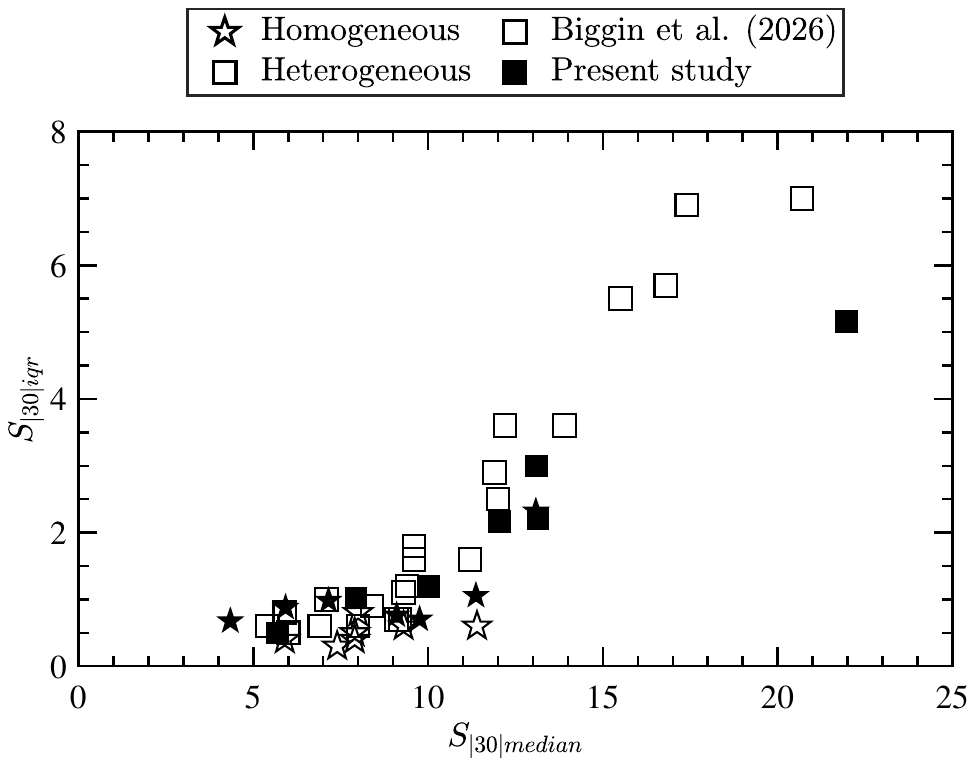}
\end{overpic}
\caption{Angular dispersion of virtual geomagnetic poles at low latitudes represented by $\Siqr$ and $\Smed$ in the present study, compared to \citet{biggin_2026}.}\label{fig:s30}
\vspace{2mm}
\end{figure}

\begin{figure}[h!]
\centering
\begin{overpic}[width=0.8\linewidth,trim={0cm 3cm 0cm 0cm},clip]{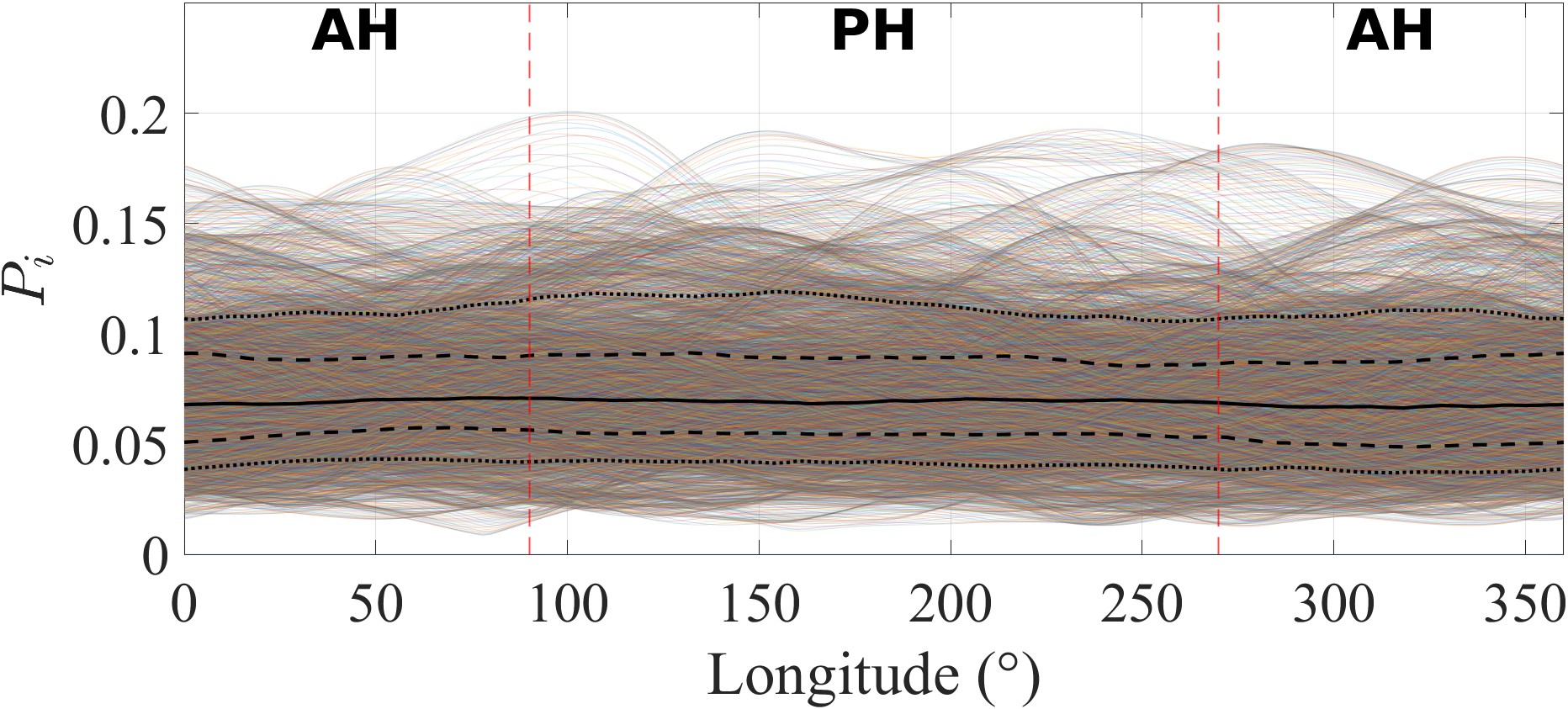}
\put(-2,45){$(a)\,\qstar=0,\ \tRaC=10^4$}
\end{overpic}

\vspace{10mm}

\begin{overpic}[width=0.8\linewidth,trim={0cm 3cm 0cm 0cm},clip]{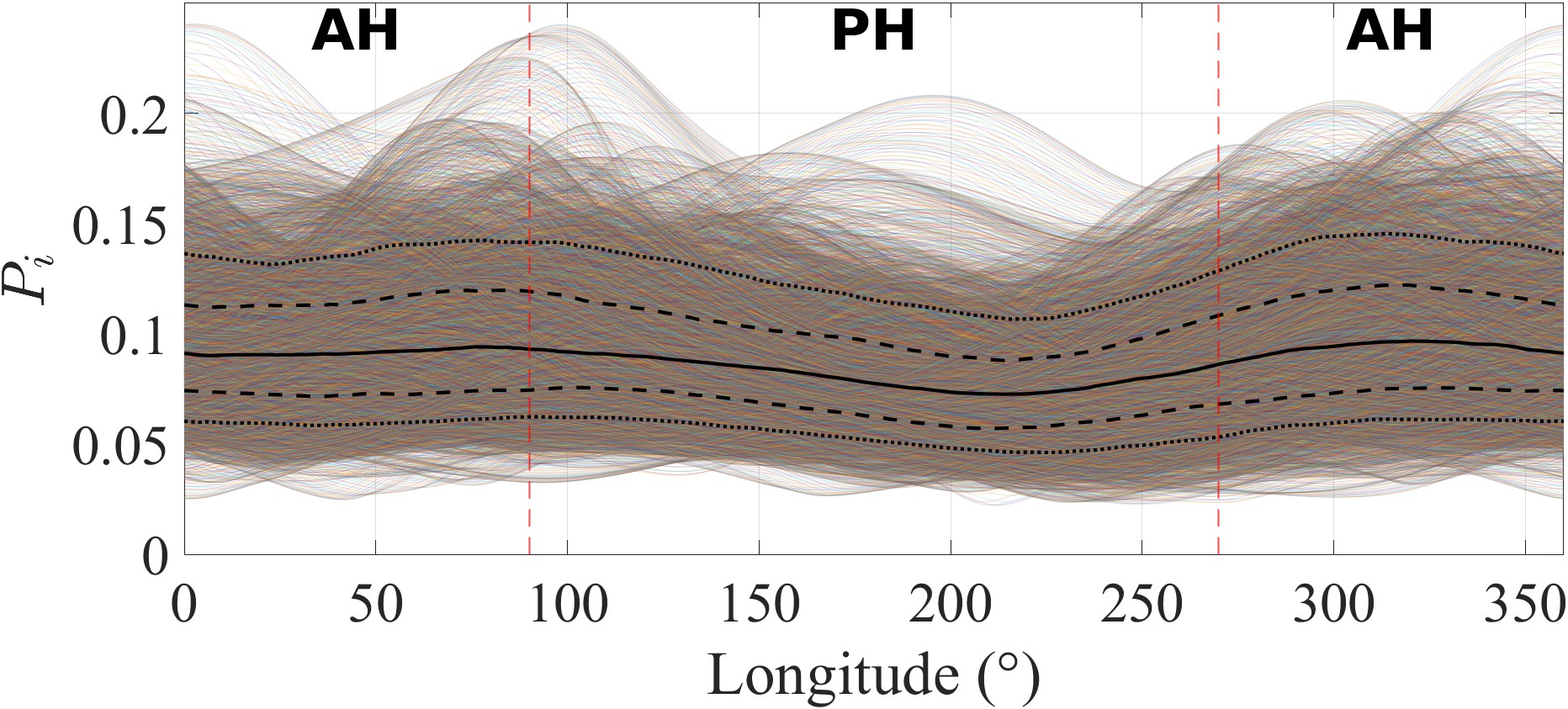}
\put(-2,45){$(b)\,\qstar=5,\ \tRaC=10^4$}
\end{overpic}

\vspace{10mm}

\begin{overpic}[width=0.8\linewidth,trim={0cm 0cm 0cm 0cm},clip] {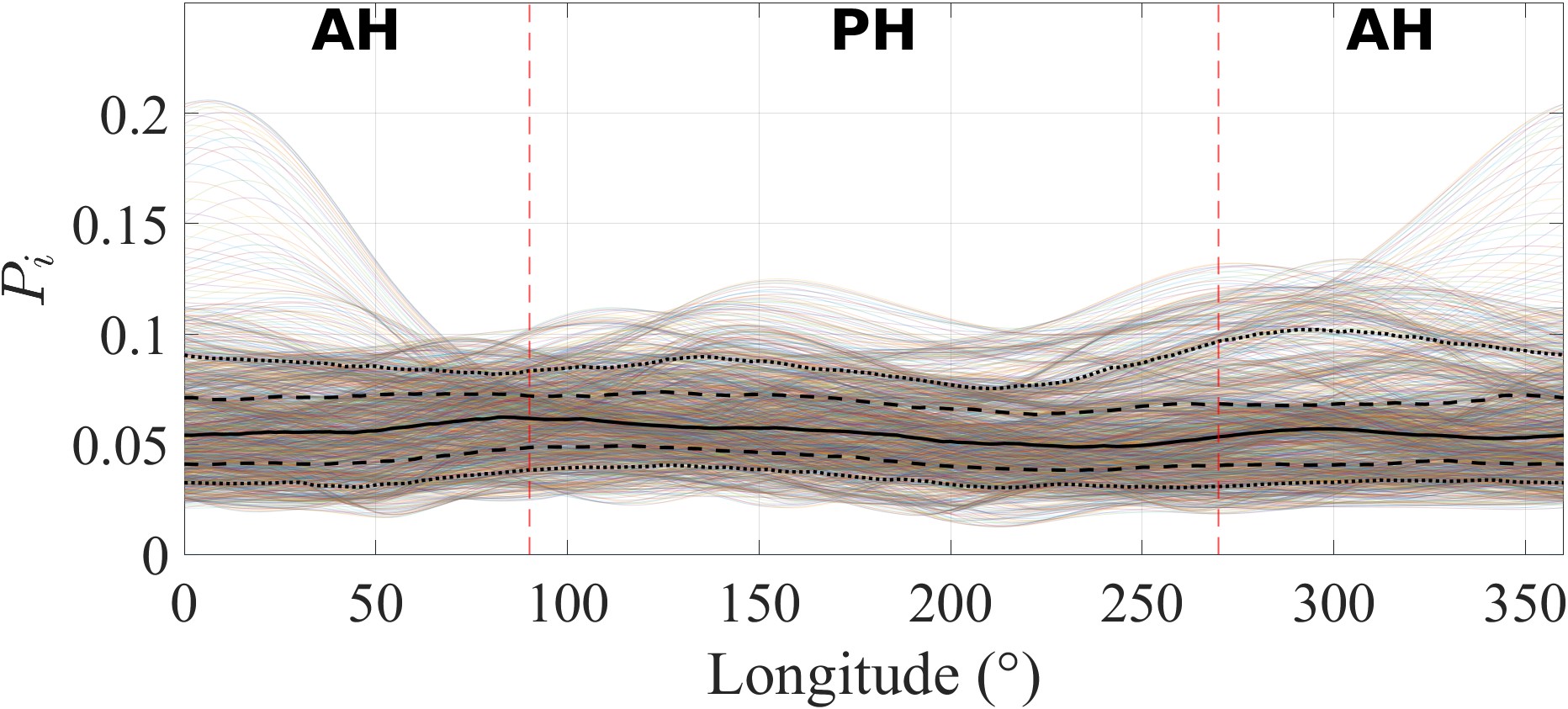}
\put(-2,50){$(c)\,\qstar=5,\ \tRaC=10^5$}
\end{overpic}
\caption{Longitudinal distribution of the latitudinally-averaged paleo-activity index $P_i$ for each time step (thin coloured lines), median in time (thick solid line). The $25^{th}-75^{th}$ and $10^{th}-90^{th}$ inter-quartile ranges are also marked by dashed and dotted lines, respectively. The surface magnetic field is truncated at $\truncation=5$ for the three simulations (a) homogeneous, $\tRaC = 10^{4}$ (b) heterogeneous, $\tRaC = 10^{4}$ and (c) heterogeneous, $\tRaC = 10^{5}$ at $\tRaT = 550$.}\label{fig:pi_vs_lon}

\end{figure}

 We also look for persistent regional features of the magnetic field induced by the thermal RILs near the equator or the light element accumulation near the poles. These can be expressed as a departure from the GAD field in the long-term behaviour of the Earth. The following seven measures are used for this purpose.

\begin{itemize}
    \item The dipole-dominance and longitudinal variation of the TAF at Earth's surface are quantified using the ratios $\ADNADTAF$ and $\ZNDNDTAF$, calculated from the Lowes power spectra for the magnetic energy \citep{lowes_1974}, following \citet{biggin_2026}, with Gauss coefficients obtained from the simulated fields truncated at spherical harmonic degree and order $\truncation=4$. These measures are listed in \ref{app:diagnostics}.
    
    \item The polar minima in the TAF (Figure \ref{fig:pol_min}) is quantified by the time-averaged radial field magnitude ($\truncation=12$) at the north pole, $B_r(\theta = 0^\circ)$, normalised by its maximum absolute value in the northern hemisphere, $max\{|B_r(\theta\leq90^\circ)|\}$. This measure differs from previous studies \citep{cao_2018,lezin_2023}, which instead considers statistics of the polar minima derived from the instantaneous field. 
    
    \item Deviation from GAD is directly measured by the ratios of specific Gauss coefficients relative to the axial dipole (e.g., $\GC{1}{1}$ and $\GC{3}{0}$). The mean and standard deviations, as shown in Figure \ref{fig:gcratio}, are calculated from the time series of the ratios of the Gauss coefficients.
    
    \item Eccentricity of the dipole, calculated as the distance of the intersection point of the dipole at the equatorial plane from the centre of Earth \citep{james_1967,olson_2012}. The mean location of the dipole ($\dipecc$) and its standard deviation are calculated from the time series of the eccentricity as shown in Figure \ref{fig:ecc_dipole} and Figure \ref{fig:eccentricity}.
    
    \item  Longitudinal structures near the equator \citep{mound_2023} and their variability is quantified using the median ($\Smed$) and inter-quartile range ($\Siqr$) of dispersion of virtual geomagnetic poles (VGPs) at low latitudes (i.e. magnetic latitude $\maglat < 30^\circ$) with the surface field truncated at $\truncation=10$, following \citet{biggin_2026}. The values are compared with \citet{biggin_2026} in Figure \ref{fig:s30}. 
    
    \item Longitudinal variation of the field is measured by the time- and latitudinally-averaged paleo-secular variation index $P_i$ (Figure \ref{fig:pi_vs_lon}), where the surface field is truncated at $\truncation=5$, following \citep{mason_2024}.
    
    \item Another measure, is the magnitude and spatial variability of the inclination anomaly ($\IncAnom = \Inc - \IncD$), where $\Inc$ is the field inclination at a location ($\theta,\phi$) on Earth's surface and $\IncD$ is the expected dipole inclination \citep{butler_2004}. The longitudinal variability of inclination anomaly is calculated from the time-averaged $\IncAnom$ map at Earth's surface (Figure \ref{fig:inc_anom_map}). Here, at each latitude, we define the range of $\IncAnom$ as the difference between the maximum and minimum $\IncAnom$ in the longitudinal distribution (Figure \ref{fig:inc_anom_range}).
\end{itemize}

\clearpage

\subsection{Non-zonal features of the time-averaged field}\label{sec:nonzon_TAF}
\begin{figure}[h!]
\centering
\raisebox{-0.5cm}{\begin{overpic}[width=0.345\linewidth,trim={4cm 2cm 3.2cm 2.9cm},clip]{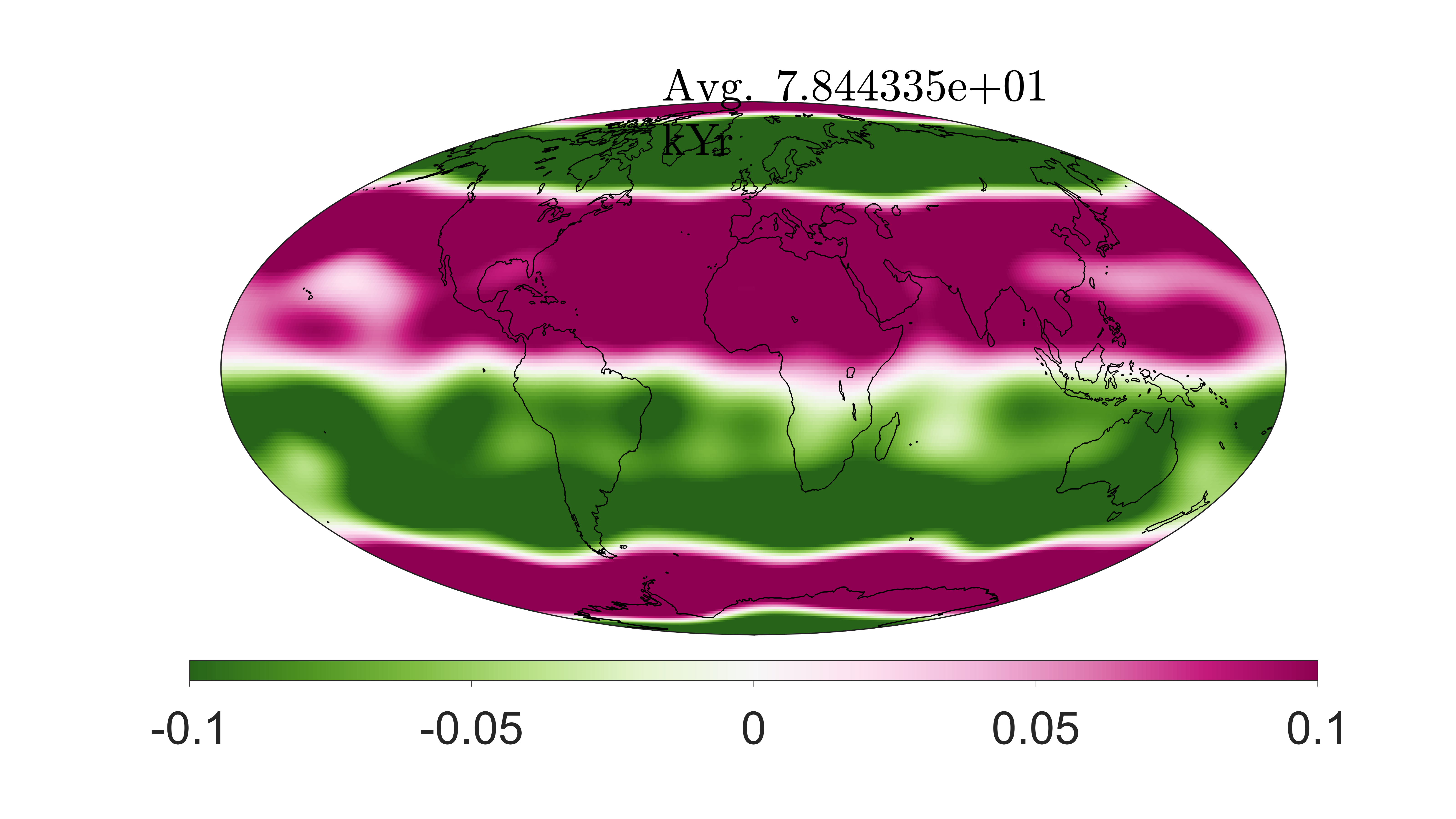}
\put(-2,60){$(a)\tRaC=10^{4},\ \qstar=0$}
\end{overpic}}
\begin{overpic}[width=0.3\linewidth,trim={0cm 0cm 8cm 1.1cm},clip]{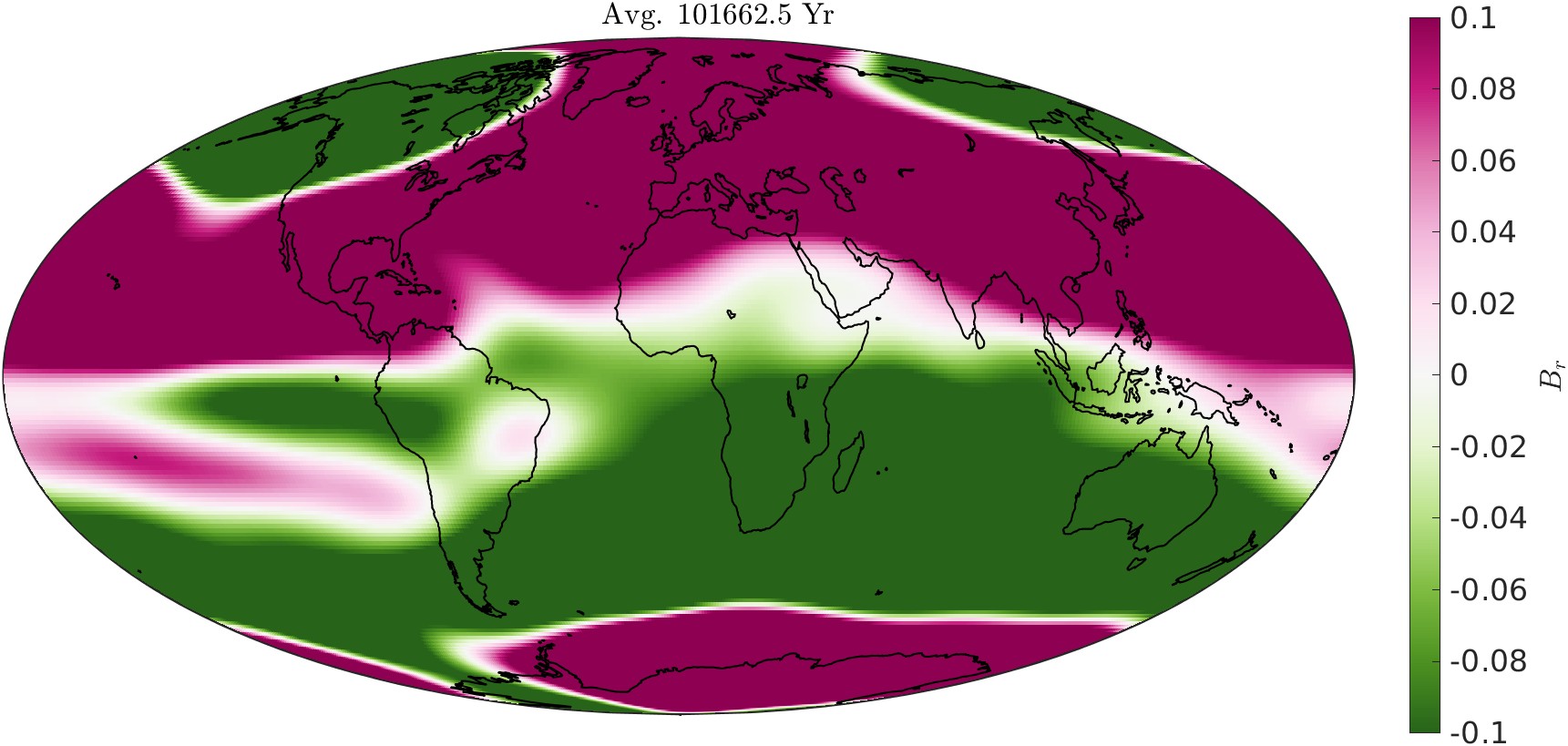}
\put(-2,59){$(b)\tRaC=10^{4},\ \qstar=5$}
\end{overpic}
\hspace{1mm}
\begin{overpic}[width=0.3\linewidth,trim={0cm 0cm 8cm 1.1cm},clip]{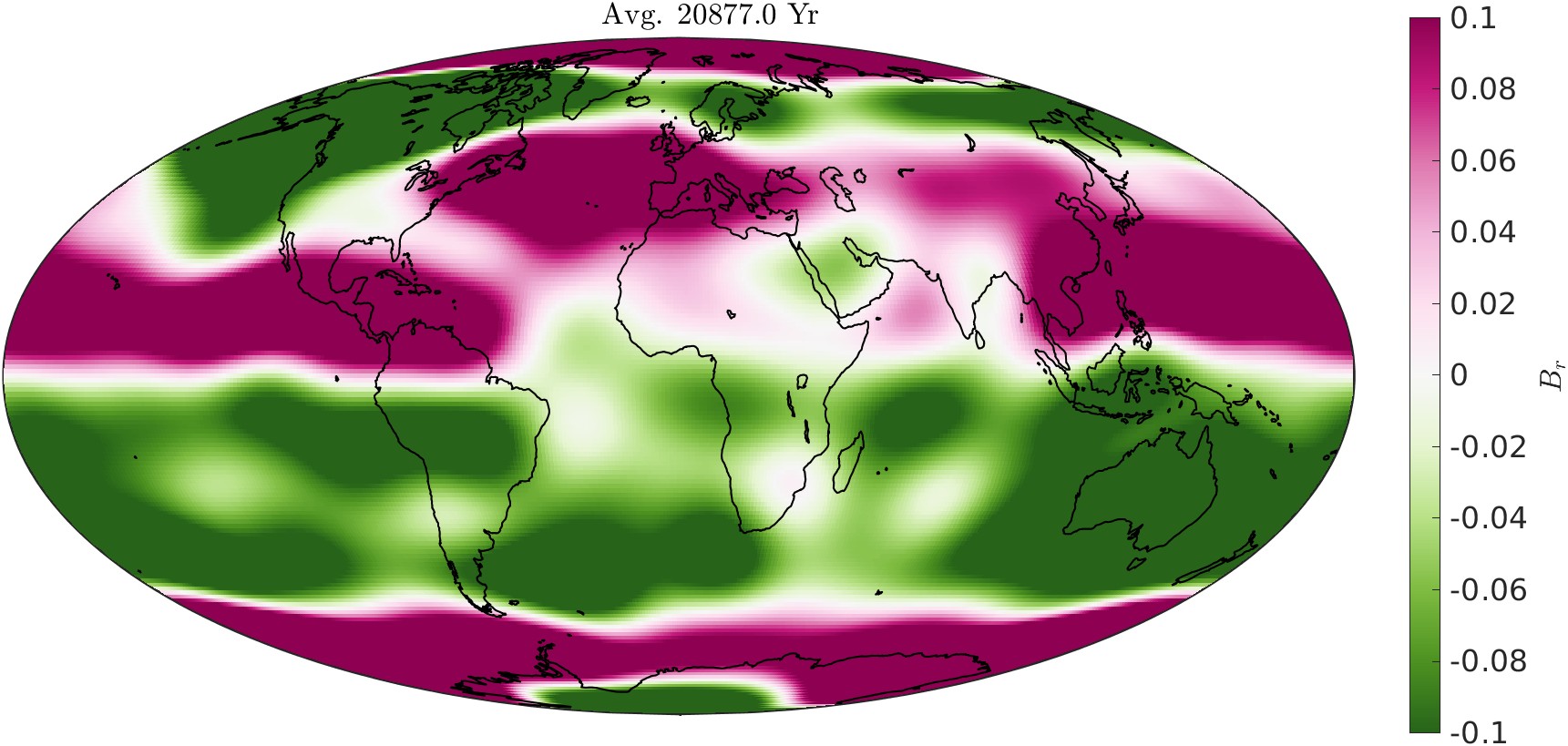}
\put(-2,59){$(c)\tRaC=10^{5},\ \qstar=5$}
\end{overpic}
\caption{Time-averaged non-dipolar ($\ell\ge2$) radial magnetic field at the CMB truncated at $\truncation=12$.}\label{fig:br_ndip}
\vspace{2mm}
\end{figure}

\begin{figure}[h!]
\begin{overpic}[width=\linewidth,trim={0cm 11.7cm 0cm 10cm},clip]{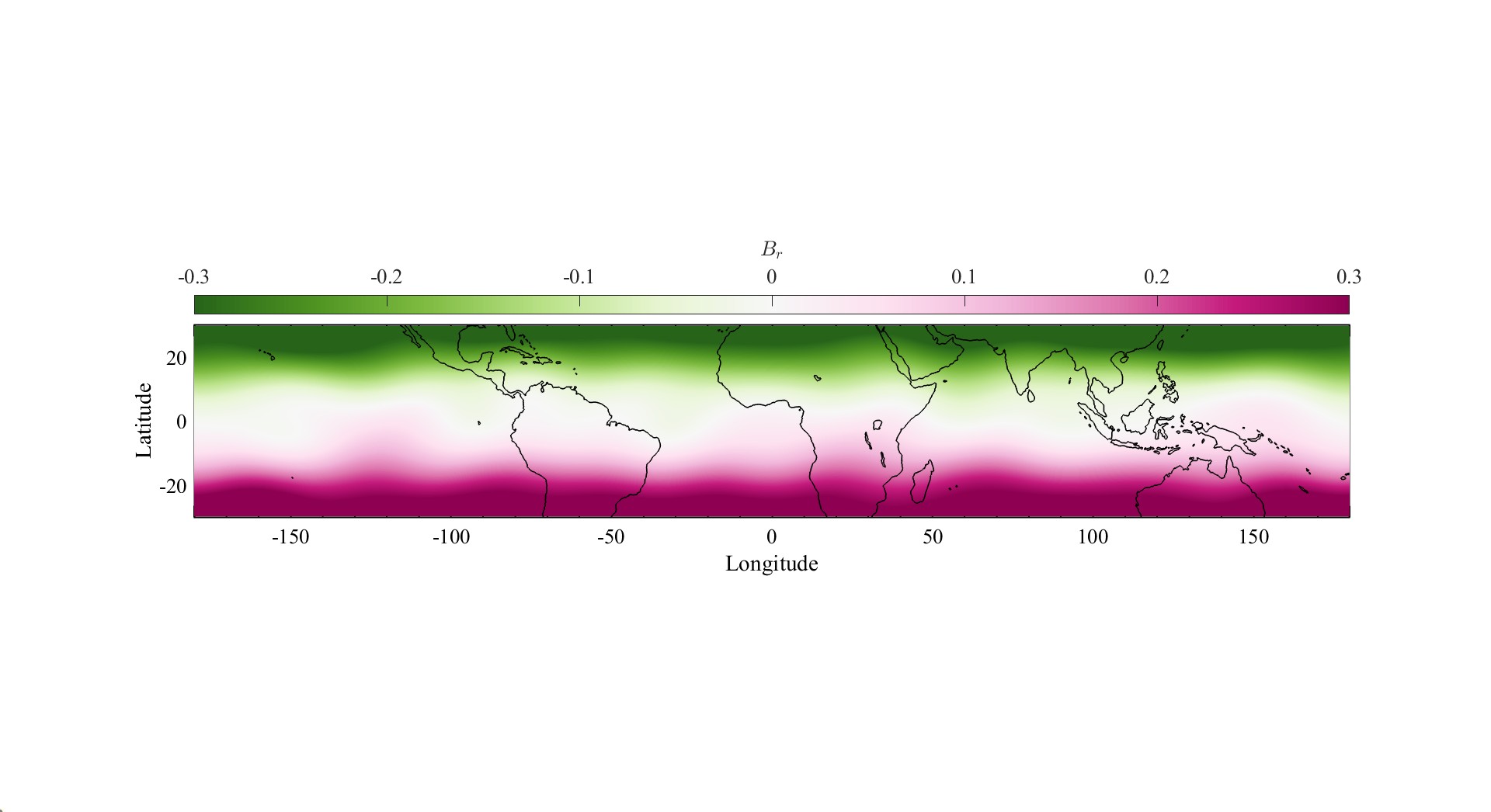}
\put(5,20){$(a)$}
\end{overpic}

\vspace{5mm}

\begin{overpic}[width=\linewidth,trim={0cm 11.7cm 0cm 14.7cm},clip]{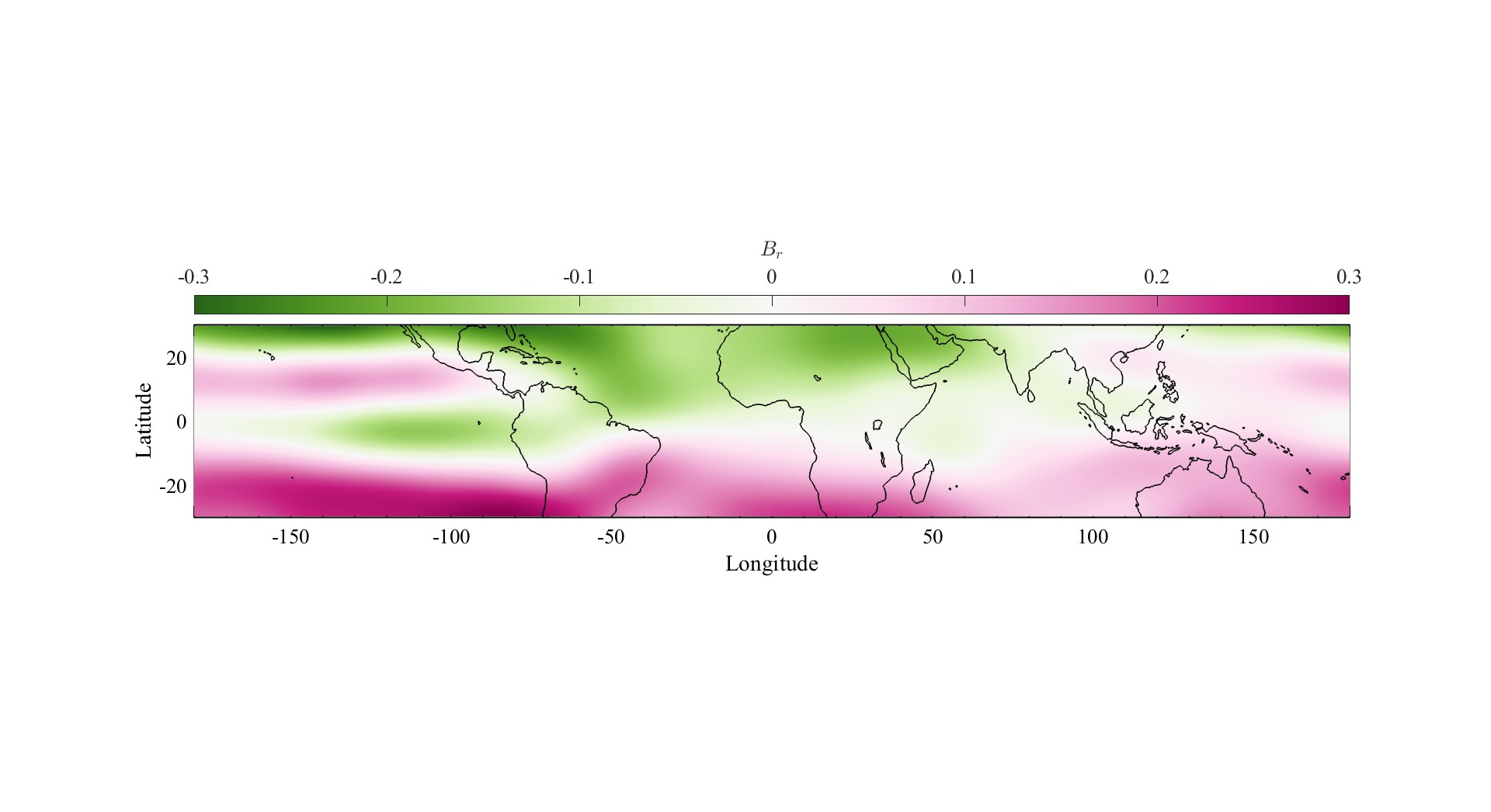}
\put(5,18){$(b)$}
\end{overpic}

\vspace{5mm}

\begin{overpic}[width=\linewidth,trim={0cm 10cm 0cm 14cm},clip]{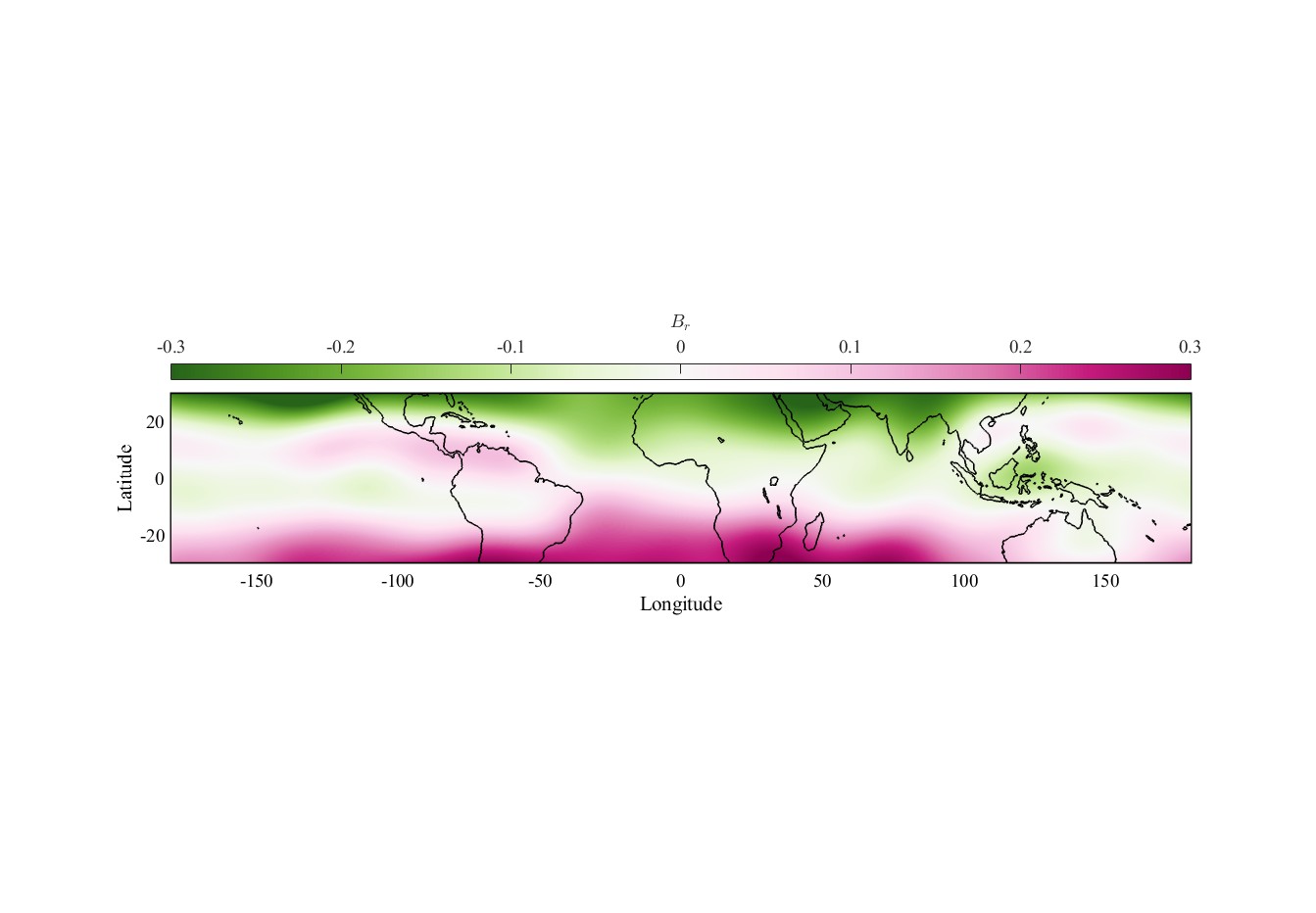}
\put(5,20){$(c)$}
\end{overpic}
\caption{Time-averaged radial magnetic field at the CMB near the equatorial region (i.e. $-30^\circ$ degree to $30^\circ$ latitude) truncated at $\truncation=12$ for (a) $\qstar=0$, $\tRaC=10^{4}$ (b) $\qstar=5$, $\tRaC=10^{4}$, (c) $\qstar=5$, $\tRaC=10^{5}$.  }\label{fig:Br_eq}
\end{figure}

Non-zonal features in the non-dipolar part of the field (Figure \ref{fig:br_ndip}) and the total radial field at low latitudes (Figure \ref{fig:Br_eq}). 

\vspace{5mm}

\clearpage

\subsection{Properties of the stratified regions}\label{sec:thk_RILs}
\begin{figure}[h!]
\centering
\vspace{5mm}
\begin{overpic}[width=0.9\linewidth,trim={0cm 0cm 0cm 0cm},clip]{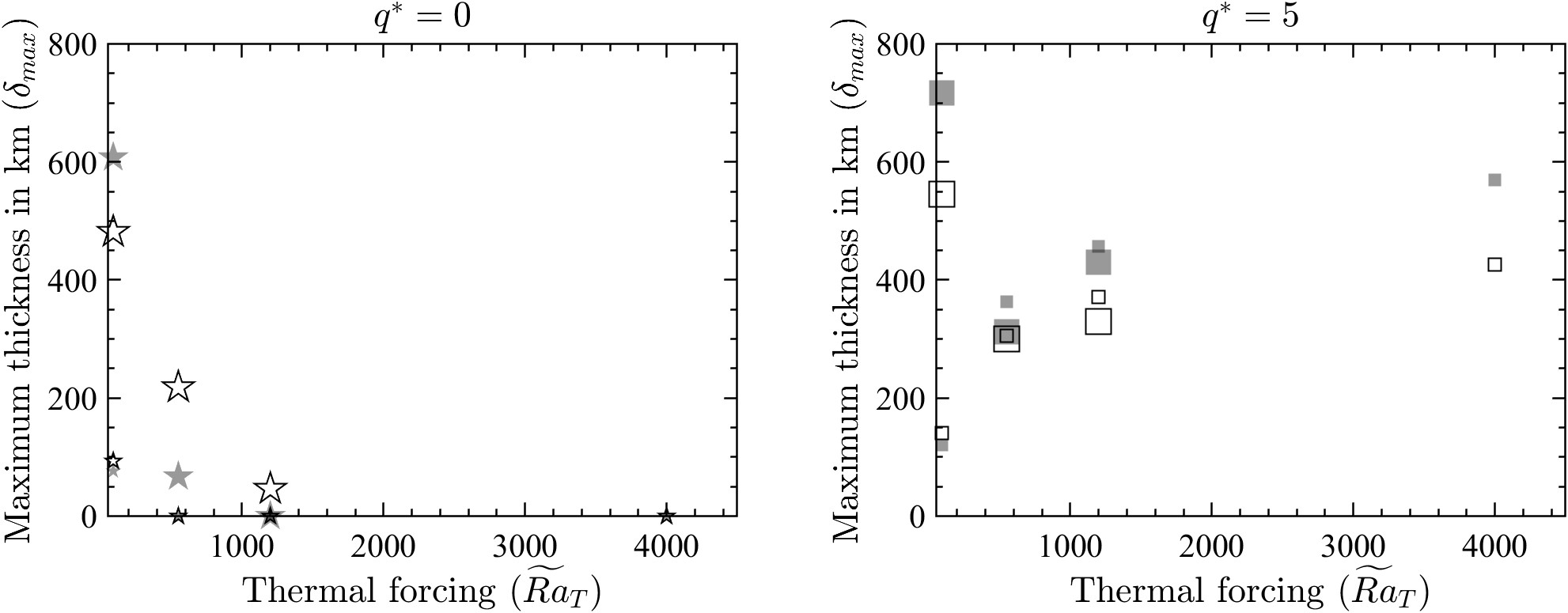}
\put(-4,40){$(a)$}
\put(48,40){$(b)$}
\end{overpic}

\vspace{5mm}

\begin{overpic}[width=0.9\linewidth,trim={0cm 0cm 0cm 0cm},clip]{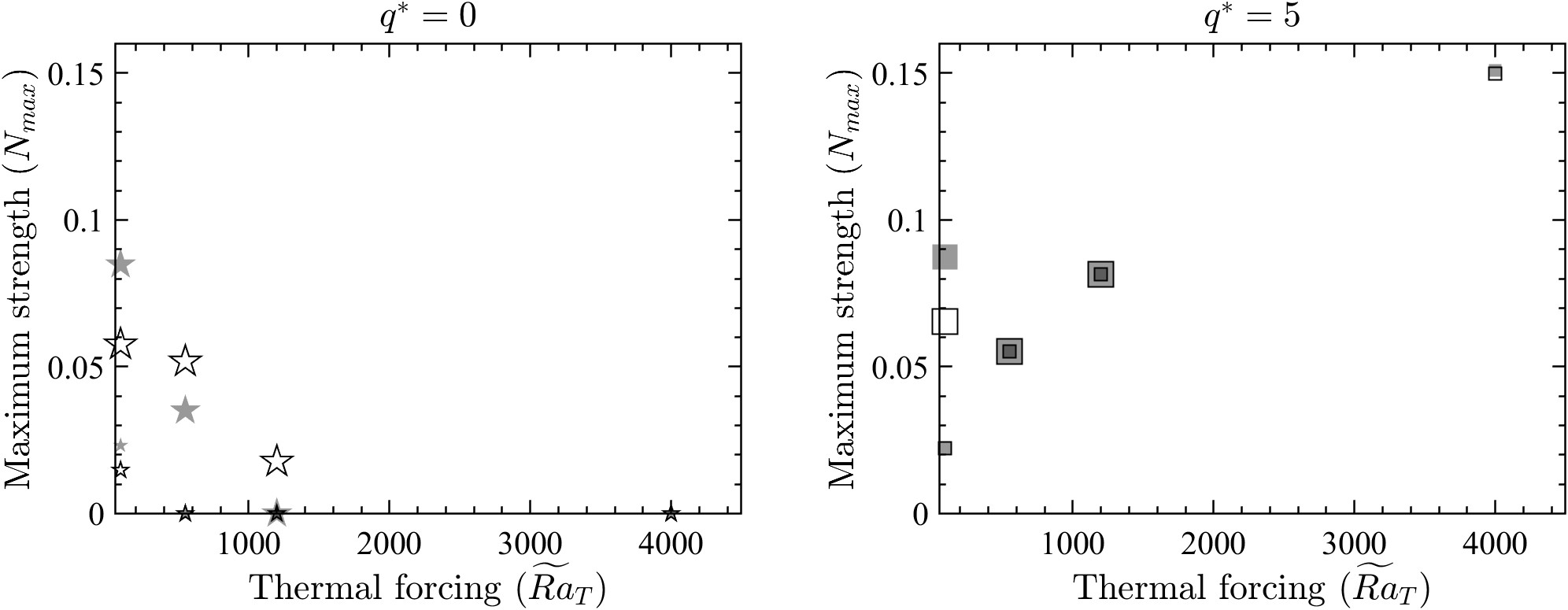}
\put(-4,40){$(c)$}
\put(48,40){$(d)$}
\end{overpic}
\caption{Variation of (a) the thickness and (b) the strength of stable regions with thermal and chemical forcing. Results from the present thermochemical geodynamo models (filled symbols) are compared with convection simulations of \citet{naskar_2026} at identical parameter values.}\label{fig:str_thk_RIL}. 
\end{figure}

The strength and thickness of the stratified regions in our dynamo simulations are compared with the corresponding non-magnetic convection simulations \citep{naskar_2026} in Figure \ref{fig:str_thk_RIL}.

\clearpage

